\documentclass[a4paper,fleqn,usenatbib]{mnras}
\usepackage{mathptmx}
\usepackage[T1]{fontenc}
\usepackage{ae,aecompl}
\usepackage{times}
\usepackage{amssymb}
\usepackage{amsmath}
\usepackage{upgreek}
\usepackage{color}
\usepackage{graphicx}
\usepackage{pdflscape}

\newcommand{\msun}{\mbox{M$_\odot$}}% Msun
\newcommand{\mean}[1]{\mbox{$\langle#1\rangle$}} %generic mean for defined qu.

\title[ATOMS II]{ATOMS: ALMA Three-millimeter Observations of Massive Star-forming regions -- II. Compact objects in ACA observations and star formation scaling relations}

\author[T. Liu et al.]{
Tie Liu,\thanks{E-mail: liutie@shao.ac.cn (TL)}$^{1,2}$
Neal J. Evans,$^{3,2}$
Kee-Tae Kim,$^{2,4}$
Paul F. Goldsmith,$^{5}$
Sheng-Yuan Liu,$^{6}$
\newauthor
Qizhou Zhang,$^{7}$
Ken'ichi Tatematsu,$^{8}$
Ke Wang,$^{9}$
Mika Juvela,$^{10}$
Leonardo Bronfman,$^{11}$
\newauthor
Maria. R. Cunningham,$^{12}$
Guido Garay,$^{11}$
Tomoya Hirota,$^{8}$
Jeong-Eun Lee,$^{13}$
Sung-Ju Kang,$^{2}$
\newauthor
Di Li,$^{14,15}$
Pak-Shing Li,$^{16}$
Diego Mardones,$^{11}$
Sheng-Li Qin,$^{17}$
Isabelle Ristorcelli,$^{18}$
\newauthor
Anandmayee Tej,$^{19}$
L. Viktor Toth,$^{20}$
Jing-Wen Wu,$^{14}$
Yue-Fang Wu,$^{21}$
Hee-weon Yi,$^{13}$
\newauthor
Hyeong-Sik Yun,$^{13}$
Hong-Li Liu,$^{22}$
Ya-Ping Peng,$^{23}$
Juan Li,$^{24,25}$
Shang-Huo Li,$^{24}$
\newauthor
Chang-Won Lee,$^{2,4}$
Zhi-Qiang Shen,$^{24,25}$
Tapas Baug,$^{9}$
Jun-Zhi Wang,$^{24,25}$
Yong Zhang,$^{26}$
\newauthor
Namitha Issac,$^{19}$
Feng-Yao Zhu,$^{1}$
Qiu-Yi Luo,$^{1}$
Xun-Chuan Liu,$^{21}$
Feng-Wei Xu,$^{21}$
\newauthor
Yu Wang,$^{21}$
Chao Zhang,$^{17}$
Zhiyuan Ren,$^{14}$
Chao Zhang$^{14}$
\\
Affiliations are listed at the end of the paper}
\date{Accepted XXX. Received YYY; in original form ZZZ}

\pubyear{2020}

\begin{document}
\label{firstpage}
\pagerange{\pageref{firstpage}--\pageref{lastpage}}
\maketitle

\begin{abstract}

We report studies of the relationships between the total bolometric luminosity ($L_{\rm bol}$ or $L_{\rm TIR}$)
and the molecular line luminosities of $J=1-0$ transitions of H$^{13}$CN, H$^{13}$CO$^+$, HCN, and HCO$^+$ with data obtained from ACA observations in the "ATOMS" survey of 146 active Galactic star forming regions.  The correlations between $L_{\rm bol}$ and molecular line luminosities $L'_{\rm mol}$ of the four transitions all appear to be approximately linear. Line emission of isotopologues shows as large scatters in $L_{\rm bol}$-$L'_{\rm mol}$ relations as their main line emission. The log($L_{\rm bol}$/$L'_{\rm mol}$) for different molecular line tracers have similar distributions. The $L_{\rm bol}$-to-$L'_{\rm mol}$ ratios do not change with galactocentric distances ($R_{\rm GC}$) and clump masses ($M_{\rm clump}$). The molecular line luminosity ratios (HCN-to-HCO$^+$, H$^{13}$CN-to-H$^{13}$CO$^+$, HCN-to-H$^{13}$CN and HCO$^+$-to-H$^{13}$CO$^+$) all appear constant against $L_{\rm bol}$, dust temperature ($T_{\rm d}$),  $M_{\rm clump}$ and $R_{\rm GC}$.  Our studies suggest that both the main lines and isotopologue lines are good tracers of the total masses of dense gas in Galactic molecular clumps. The large optical depths of main lines do not affect the interpretation of the slopes in star formation relations. We find that the mean star formation efficiency (SFE) of massive Galactic clumps in the "ATOMS" survey is reasonably consistent with other measures of the SFE for dense gas, even those using very different tracers or examining very different spatial scales.

\end{abstract}


%%%%%%%%%%%%%%%%%%%%%%%%%%%%%%%%%%%%%%%%%%%%%%%%%%

%%%%%%%%%%%%%%%%% BODY OF PAPER %%%%%%%%%%%%%%%%%%

\section{Introduction}

Since the pioneering works by \citet{Gao2004} and \citet{Wu2005}, many observational studies toward external galaxies \citep{Garcia2006,Juneau2009,Garcia2012,Greve2014,Zhang14,Liu2015,Chen2015,Tan2018,Jimenez2019} and Galactic molecular clouds \citep{Wu2010,Liu2016,Stephens2016} have revealed a strong linear relationship between the recent star formation rate (SFR), as traced by the total infrared emission, and the dense molecular gas mass that is indicated by line luminosities ($L'_{\rm mol}$) of dense molecular gas tracers (e.g., HCN, HCO$^{+}$, and CS). This so called "dense gas star formation law" may imply that the dense molecular gas rather than the total molecular gas is the direct fuel for star formation \citep{Kennicutt2012,Vutisalchavakul2016}. While significant progress has been made in recent years, the origin of this relationship is still under debate.

The $J=1-0$ transitions of HCN and HCO$^+$ are among the most commonly used tracers in the studies of "dense gas star formation law". However, the emission lines of these two dense gas tracers tend to be optically thick in dense parts of molecular clouds \citep[e.g.,][]{Sanhueza2012,Hoq2013,Shimajiri2017}. In the limit that all transitions are optically thin and only collisional excitation is important, but densities are far below the critical density, the emission would be proportional to $n \times n_{\rm mol}$, where $n$ is the number density of colliders (mainly H$_2$) and $n_{\rm mol}$ is the density of the line-emitting molecules, thus proportional to $n^2$ for constant abundance. At the other extreme, the emission from an optically thick, thermalized line (e.g. $^{12}$CO) is not sensitive to density at all. While the extremes rarely apply, and the rich energy level structure of molecules permits more interesting excitation (masers, cooling of excitation temperatures below the temperature of the CMB, etc.), the above limiting cases illustrate why  the emission from rare isotopologues is more highly weighted toward denser regions than their corresponding main lines. To capture some of the effects, the concept of effective excitation density was introduced to indicate the density needed to produce a line of a characteristic strength (1 K km~s$^{-1}$), given a column density typical for each species \citep{Evans1999,Shirley2015}. The low effective excitation densities of the two main transitions of HCN and HCO$^+$ \citep[4.5$\times10^3$ cm$^{-3}$ for the HCN $J=1-0$ line and 5.3$\times10^2$ cm$^{-3}$ for the HCO$^+$ $J=1-0$ line at 20 K;][]{Shirley2015} indicate that they can be excited in low density ($n<10^3$ cm$^{-3}$) gas and are often optically thick.

Indeed, recent observations of nearby Giant molecular clouds \citep{Kauffmann2017,Pety2017,Shimajiri2017} indicate that HCN (1-0) and HCO$^+$ (1-0) are easily detected in extended translucent regions at a typical density of 500 cm$^{-3}$ and are poor tracers of dense structures such as filaments or cores. \citet{Stephens2016} have argued that most of the Galaxy's luminosity of HCN may arise from distributed, sub-thermal emission rather than from dense gas. Most recently, \citet{Evans2020} also found that a substantial fraction (in most cases, the majority) of the total HCN (1-0) and HCO$^+$ (1-0) line luminosity in six distant ($d\sim$3.5-10.4 kpc) clouds arises in gas below the $A_{\rm V}\sim$8 mag threshold, above which the vast majority of dense cores and YSOs are found in nearby clouds \citep{Heiderman2010,Lada2010,Lada2012}. These studies have challenged the ideas that these commonly used tracers (e.g., $J=1-0$ of HCN and HCO$^+$) can reveal well the spatial distribution of star forming gas in clouds.

As discussed above, in contrast to the main lines, emission from the isotopologues are optically thinner because of their much lower abundances. The effective  densities of H$^{13}$CN $J=1-0$ and H$^{13}$CO$^+$ $J=1-0$ at 20 K are about 1.6$\times10^5$ cm$^{-3}$ and 2.2$\times10^4$ cm$^{-3}$, respectively, which are about 40 times higher than their main lines \citep{Shirley2015}. Therefore, they can potentially be better tracers of the column densities and dense structures in molecular clouds \citep[e.g.,][]{Pety2017,Shimajiri2017}. The observations of these isotopologues can also help estimate the opacity of their corresponding main line transitions, enabling us to quantify how opacity affects the "dense gas star formation law" with main lines. Studies of the "dense gas star formation law'' with isotopologues, however, are very rare. \citet{Stephens2016} used H$^{13}$CO$^+$ $J=1-0$ in their studies but they detected the H$^{13}$CO$^+$ $J=1-0$ emission only in dozens of sources in their sample of $\sim$300 clumps. In addition, most stars form in gravitationally bound structures (clumps or cores) in molecular clouds. Therefore, it is also essential to evaluate how those lines (main lines and their isotopologues) trace the total gas mass of these gravitationally bound structures.

In this work, we investigate the "dense gas star formation law" with H$^{13}$CN $J=1-0$ and H$^{13}$CO$^+$ $J=1-0$ lines as well as their corresponding main lines toward a large sample of 146 Galactic star forming clumps. By doing this, we evaluate the reliability of the main lines being used to trace the total dense gas mass.

This study is part of an ALMA survey program, the "ATOMS" (ALMA Three-millimeter Observations of Massive Star-forming regions). The sample and observations of the "ATOMS" survey are introduced in \citet{Liu2020}. The "ATOMS" observed a large sample of 146 active star forming regions with IRAS colors characteristic of UC H{\sc ii} regions \citep{Bronfman1996}. More than 90\% of the sources in the "ATOMS" sample are potentially forming high-mass stars \citep{Liu2020}. In this paper, we only use the data from observations with the Atacama Compact 7 m Array (ACA; Morita Array).  The ACA data are particularly useful for this study because they respond primarily to the scale of clumps ($\ga$0.2 pc), rather than cores ($\la$0.1 pc). The dense gas star formation relations will surely break down at the level of cores, but will plausibly be relevant on the scale of clumps.

\section{Observations}

The ALMA observations of the ``ATOMS" survey have been summarized in \citet{Liu2020}. We here briefly introduce the observations with the ACA. Observations with the ACA were conducted from late September to mid November in 2019 with band 3 (Project ID: 2019.1.00685.S; PI: Tie Liu). The IRAS names, phase centers (columns 2-3), systemic velocities (column 4), distances (column 5), and galactocentric distances (column 6) of the targeted sources are listed in Table \ref{continuum}. The typical ACA observing time per source is $\sim$ 8 minutes. The angular resolution and maximum recovered angular scale (MRS) in ACA observations are $\sim13.1\arcsec-13.8\arcsec$ and $\sim53.8\arcsec-76.2\arcsec$, respectively. The ACA observations are sensitive to angular scales smaller than $\sim60\arcsec$ but the MRS is larger than the angular sizes of most clumps in the sample \citep{Faundez2004}. The $J=1-0$ transitions of HCN, HCO$^+$, H$^{13}$CN and H$^{13}$CO$^+$ are included in four spectral windows in the lower side-band. The spectral resolution for HCN $J=1-0$ and HCO$^+$ $J=1-0$ is 61.035 kHz (or $\sim$0.2 km~s$^{-1}$), while the spectral resolution for H$^{13}$CN $J=1-0$ and H$^{13}$CO$^+$ $J=1-0$ is 0.122 MHz (or $\sim$0.4 km~s$^{-1}$).
Calibration and imaging were carried out using the CASA software package version 5.6 \citep{McMullin2007}. The continuum visibility data are constructed using the line-free spectral channels. All images are primary beam corrected.  The achieved sensitivity of the ACA observations is around 50-90 mJy~beam$^{-1}$ per 0.122 MHz channel for lines.

\section{Results}

\subsection{Compact objects}

\begin{figure*}
\centering
\includegraphics[angle=0,scale=0.75]{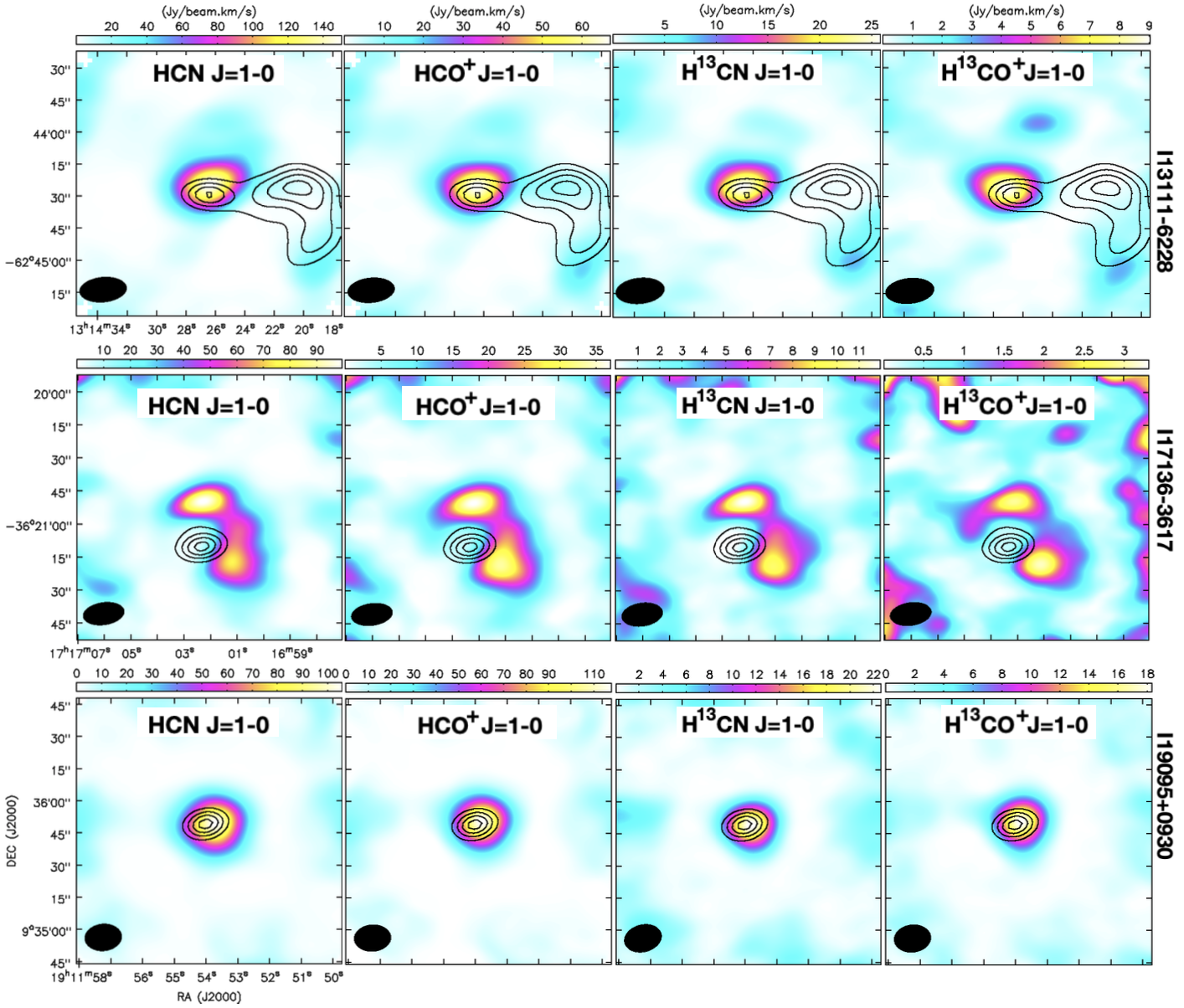}
\caption{Integrated intensity maps of four molecular lines are shown as color-scale images for three example sources. I13111-6228, I17136-3617 and I19095+0930 are shown in upper, middle and lower panels, respectively. The 3 mm continuum is shown in contours. The contour levels are from 30\% to 90\%  in steps of 20\% of peak values. The peak values of 3 mm continuum emission for I13111-6228, I17136-3617 and I19095+0930 are 0.088,  1.814 and 0.307 Jy~beam$^{-1}$, respectively.  \label{lineimage}}
\end{figure*}

We extracted compact objects from 3 mm continuum emission maps and the integrated intensity maps of the four molecular lines. Figure \ref{lineimage} shows the 3 mm continuum emission and integrated intensity maps for three sources. The compact objects in 3 mm continuum emission and molecular line emission can be easily identified by eye. In total, we detected 173, 184, 190, 189 and 182 compact sources from 3 mm continuum emission, H$^{13}$CN, H$^{13}$CO$^+$, HCN, and HCO$^+$ line emission, respectively.

As shown in Figure \ref{lineimage} for I13111-6228, we identified multiple compact objects in 3 mm continuum in 27 targets. Some targeted sources also contain multiple objects in molecular line emission as shown in Figure \ref{lineimage} for I17136-3617. We identified multiple objects in 30, 32, 35 and 27 sources from H$^{13}$CN, H$^{13}$CO$^+$, HCN, and HCO$^+$ integrated intensity maps, respectively. However, the majority ($\sim$80\%) of sources contain only a single compact object in 3 mm continuum emission and in molecular line emission as I19095+0930 in Figure \ref{lineimage}.

From 2D gaussian fits, we derived the relative coordinates (or offsets), aspect ratio values, effective radii ($R_{\rm eff}$), offsets from brightest continuum emission peaks ($d_{\rm peak}$), peak integrated intensity ($S_{\rm peak}$) and total integrated intensity ($S_{\rm total}$) for each compact object. The aspect ratio value is defined as the ratio between de-convolved FWHM major dimension ($a$) and minor dimension ($b$), and $R_{\rm eff}$ is defined as $R_{\rm eff}=\sqrt{ab}$. For compact objects identified in H$^{13}$CO$^+$ $J=1-0$ and HCO$^+$ $J=1-0$ line emission, we also derived the source-averaged velocity ($V_{\rm lsr}$) and velocity dispersion ($\sigma$) from their corresponding Moment 1 and Moment 2 maps. All of these parameters are summarized in Tables \ref{continuum} to \ref{HCN}.

\begin{figure*}
\centering
\includegraphics[angle=0,scale=0.65]{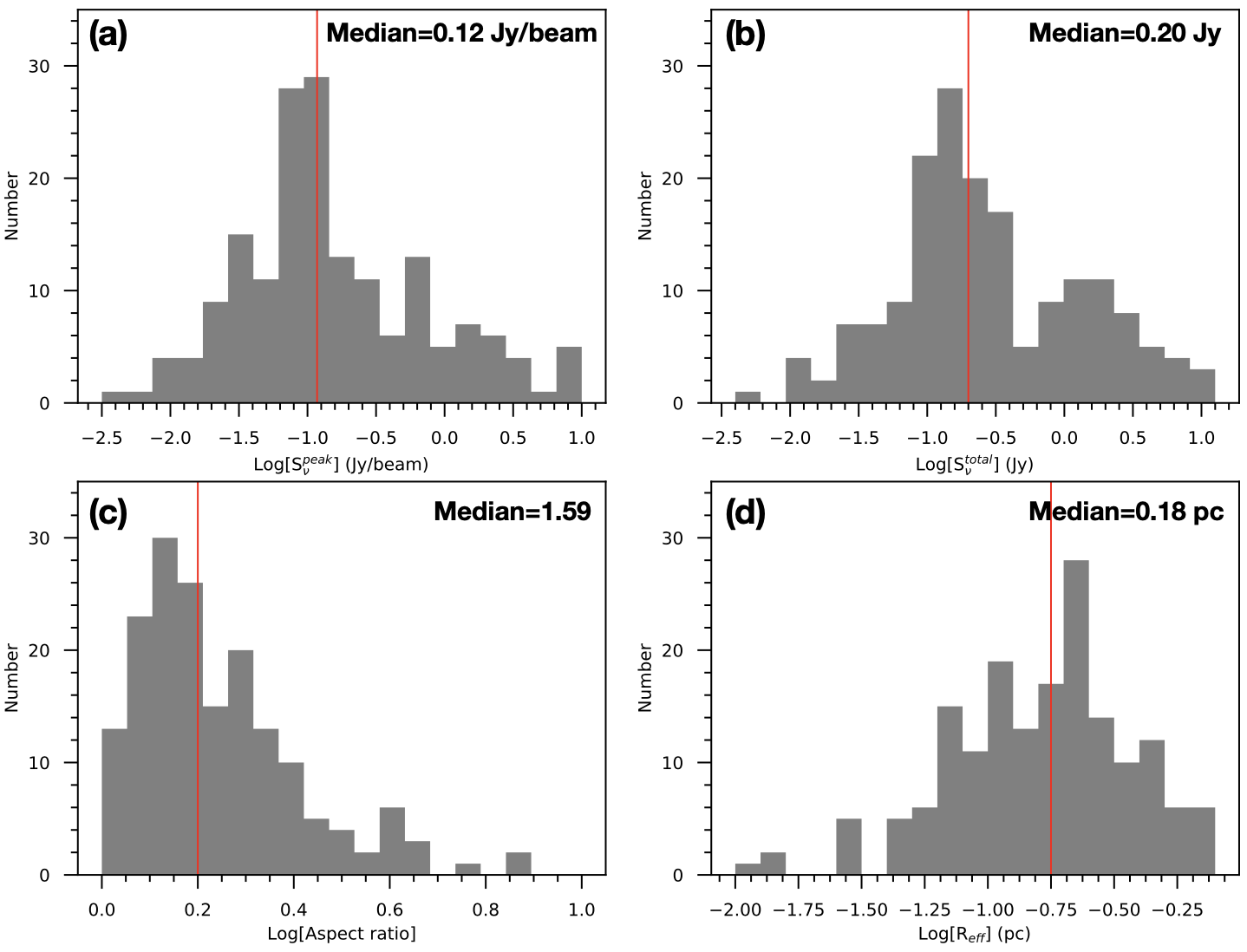}
\caption{Histograms of parameters for the compact objects identified in 3 mm continuum emission. (a) peak flux density; (b) total flux density; (c) aspect ratios; (d) effective radii. The red solid lines are median values. \label{statpara}}
\end{figure*}

\begin{figure}
\centering
\includegraphics[width=\columnwidth]{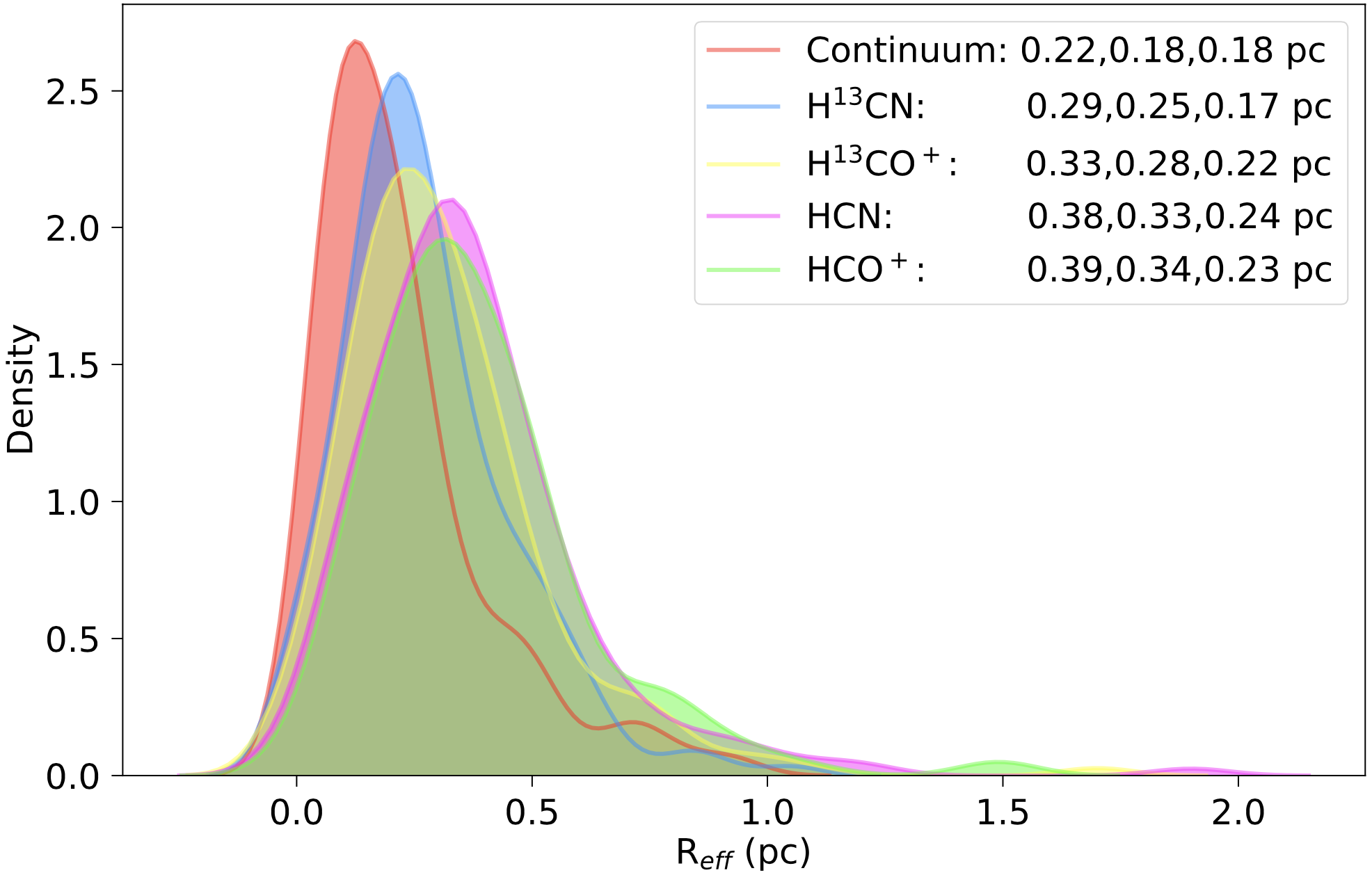}
\caption{Distribution of $R_{\rm eff}$ for compact objects identified in 3 mm continuum emission and molecular line emission, which are plotted using the function "kdeplot" in the python package Seaborn. The numbers in the upper-right box are mean, median, standard deviation values.  \label{Reff}}
\end{figure}

Figure \ref{statpara} presents histograms of parameters ($S_{\rm peak}$, $S_{\rm total}$, aspect ratio and $R_{\rm eff}$) for compact sources in 3 mm continuum emission. The median values for $S_{\rm peak}$ and $S_{\rm total}$ are 0.12 Jy~beam$^{-1}$ and 0.20 Jy, respectively. The median value of the aspect ratio is $\sim$1.6. Among the 173 compact objects identified in 3 mm continuum emission, only 18 are with aspect ratio larger than 3. The $R_{\rm eff}$ ranges from 0.01 pc to 0.94 pc with a median value of 0.18 pc. There are 45 compact objects that would be classified as dense cores with $R_{\rm eff}$ smaller than 0.1 pc. There are 14 compact objects in distant sources having $R_{\rm eff}$ larger than 0.5 pc. The detailed properties of those continuum objects will be discussed in forthcoming papers. The 3-mm continuum emission can have a significant contribution from free-free emission, compromising our ability to determine masses from the dust continuum emission. We use instead the literature values for the size (R$_{eff}$), mass (M$_{clump}$), etc. of the clumps (see Table \ref{LM}). These are taken from single-dish maps at 0.87 mm \citep{Urquhart2018} or 1.2 mm \citep{Faundez2004}, minimizing the contamination by free-free emission.

In Figure \ref{Reff}, we compare the distributions of $R_{\rm eff}$ for compact objects identified in 3 mm continuum emission and molecular line emission. The data are plotted with a gaussian kernel density estimate using the function "kdeplot" in the python package Seaborn\footnote{https://seaborn.pydata.org/generated/seaborn.kdeplot.html}. The median radii for compact objects in 3 mm continuum, H$^{13}$CN, H$^{13}$CO$^+$ HCN and HCO$^+$ line emission are 0.18, 0.25, 0.28, 0.33 and 0.34 pc, respectively. Compact objects in 3 mm continuum emission have statistically smaller $R_{\rm eff}$ than that of compact objects identified in molecular line emission. $R_{\rm eff}$ for compact objects identified in H$^{13}$CN and H$^{13}$CO$^+$ line emission are also statistically smaller than compact objects identified in HCN and HCO$^+$ line emission. Interestingly, there seems to be a trend between $R_{\rm eff}$ and effective excitation density for the lines. The higher the effective excitation density, the smaller the $R_{\rm eff}$. These results are consistent with the idea that lines with higher effective excitation density trace denser and more compact regions of molecular clouds.

In Figure \ref{cmp}a, we compare the effective radii ($R_{\rm ACA}$) of compact sources in ACA 3 mm continuum emission with the effective radii ($R_{\rm SD}$) of their natal clumps derived in single dish observations. $R_{\rm ACA}$ is linearly correlated with $R_{\rm SD}$ as expected. Since the ACA is resolving out the more extended emission, $R_{\rm ACA}$ is systematically smaller than $R_{\rm SD}$. As shown in Figure \ref{cmp}b, the velocities derived from H$^{13}$CO$^+$ $J=1-0$ line emission for compact sources are very consistent with the systemic velocities of their natal clumps from single dish CS J=2-1 line observations \citep{Bronfman1996}.

In some sources such as I17136-3617 shown in Figure \ref{lineimage}, gas emission peaks are clearly offset from 3 mm continuum emission peaks by one beam size. These sources may be highly evolved H{\sc ii} regions that have dispersed their natal gas clumps. In further analysis, we only focus on 119 sources, for which their gas emission and 3 mm continuum emission coincide well with each other (with separation smaller than one beam size of $\sim15\arcsec$). This will help reduce the evolutionary effect in "dense gas star formation law" studies.

\subsection{Virial masses}

\begin{figure*}
\centering
\includegraphics[angle=0,scale=0.37]{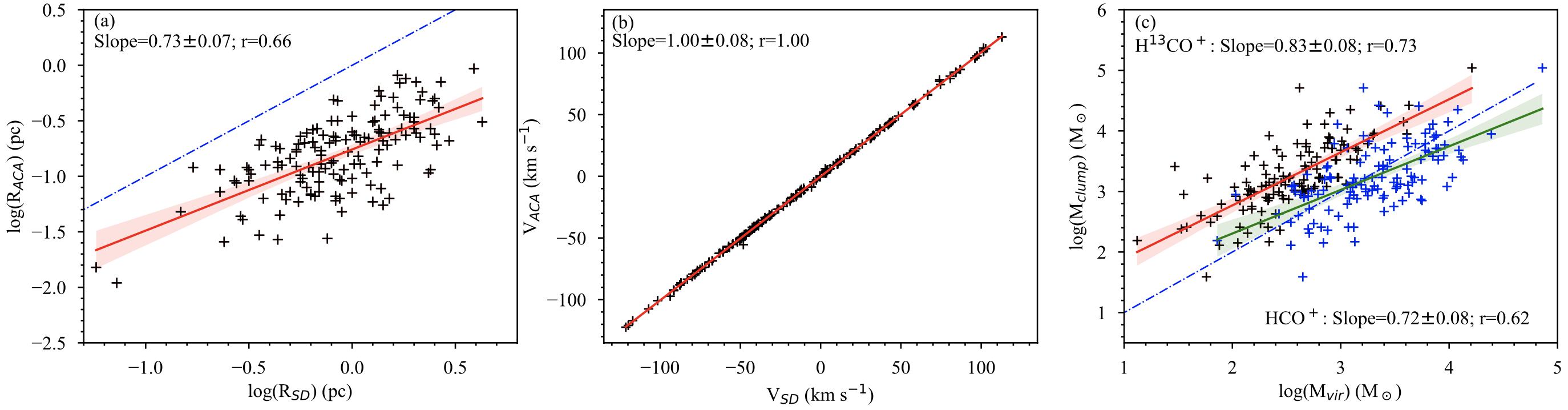}
\caption{(a) Comparison of effective radii for compact objects in ACA 3 mm continuum observations and effective radii of whole clumps in single dish observations. The blue dashed line represents that the two radii equal to each other. (b) Comparison of systemic velocities in ACA observations and single-dish observations. (c) Comparison of virial masses for compact objects identified in H$^{13}$CO$^+$ (black crosses) and HCO$^+$ (blue crosses) line emission with total clump masses from single-dish observations. The red and green lines are linear fits. The blue dashed line shows virial masses equaling to clump masses.  \label{cmp}}
\end{figure*}

We derived virial masses for compact objects identified in H$^{13}$CO$^+$  and HCO$^+$ line emission following \citet{Bertoldi1992} and \citet{Wu2010}. A correction ($a_2\sim0.9$) for a nonspherical shape in virial analysis is adopted for compact objects with aspect ratios larger than 3 \citep{Bertoldi1992}. For the others with smaller aspect ratios, this correction is negligible and has been omitted. We did not derive virial masses for compact objects identified in H$^{13}$CN or HCN because of complications caused by hyperfine structure. The virial masses for objects identified in HCO$^+$ are probably overestimated because of line broadening of HCO$^+$ resulting from the high optical depth. HCO$^+$ also covers a larger region than H$^{13}$CO$^+$ due to its lower effective excitation temperature (see Figure \ref{Reff}).

The virial masses ($M_{\rm vir}^{13}$ and $M_{\rm vir}^{12}$) derived from H$^{13}$CO$^+$ and HCO$^+$ line emission are listed in the last two columns of Table \ref{LM}. If there are more than one compact objects identified in a source, we simply sum the virial masses of the contributing objects. The virial masses derived from H$^{13}$CO$^+$ range from 13 to 16260 $M_{\sun}$ with a median value of 334 $M_{\sun}$. About 90\% of the sources have virial masses larger than 100 $M_{\sun}$. The virial masses for the main isotope are considerably larger than those based on the rare isotope. While the HCO$^+$ lines are generally optically thick, the other effect is that they extend over a larger region due to much lower effective excitation density. This leads to higher virial masses for HCO$^+$.

In Figure \ref{cmp}c, we compare the virial masses with the total clump masses from single-dish measurements \citep{Faundez2004,Urquhart2018}. The virial masses are strongly correlated with the total clump masses ($M_{\rm clump}$). The virial masses from H$^{13}$CO$^+$ are systematically smaller than the total masses of their natal clumps with a median virial parameter ($\alpha$=$M_{\rm vir}$/$M_{\rm clump}$) of $\sim$0.2. The small value probably reflects the fact that the regions probed by H$^{13}$CO$^+$ are smaller than those used to obtain the clump mass from the single-dish data. The virial masses from HCO$^+$ are comparable to the total clump masses, with a median virial parameter of $\sim$1.2, indicating that ACA observations  of HCO$^+$ trace structures similar to those traced by the single-dish millimeter continuum data, which are most likely gravitationally bound.
%Figure \ref{cmp}c may also be interpreted that the gas traced by HCO$^+$ is in a virial equilibrium, while the higher density gas traced by H$^{13}$CO$^+$ is under gravitational collapse.

Small virial parameters have also been reported in many recent measurements of the thermodynamic properties in high-mass star-forming regions \citep{Pillai2011,Kauffmann2013,Zhang2015,Hull2019}, which appear to challenge the picture of star formation in which gas evolves in a state of equilibrium \citep[e.g.,][]{Hull2019}. However, in this work and most of previous works magnetic field support were not taken into account in virial analysis. Magnetic support could be comparable to the turbulent and thermal support in high-mass star forming regions, which can bring the dense gas close to a state of equilibrium \citep{Pillai2011,Sanhueza2017,Zhang2015,Liu2018a,Liu2018b,Hull2019,Soam2019,Tang2019}. More detailed energetics comparison of the gravitational potential energy, turbulent support, thermal pressure, and magnetic support are needed in future analysis of virial equilibrium.

\subsection{The relations between infrared luminosities and molecular line luminosities}

\begin{figure*}
\centering
\includegraphics[angle=0,scale=0.45]{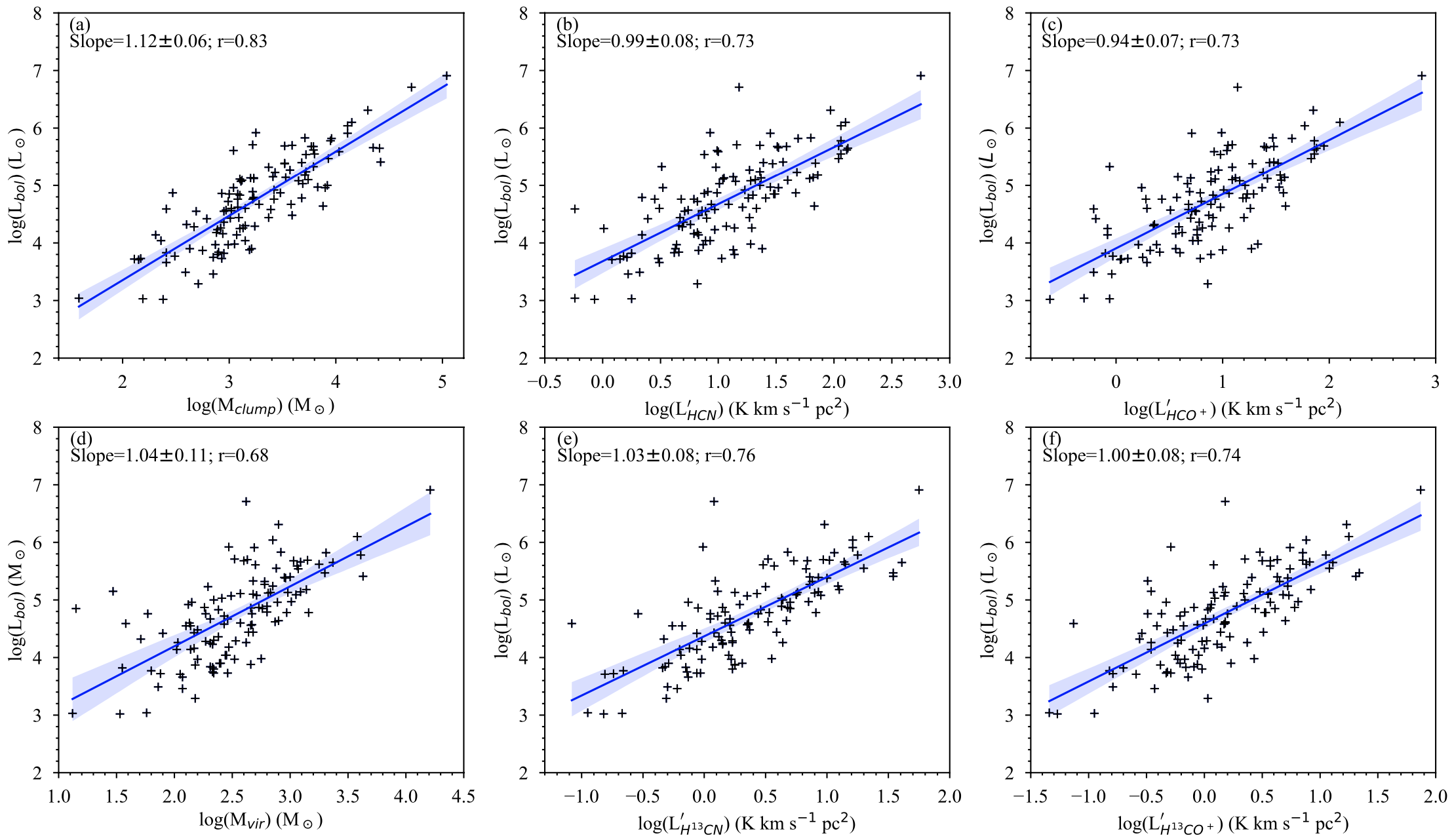}
\caption{The correlations between $L_{\rm bol}$ and other parameters ($M_{\rm clump}$, $M_{\rm vir}$ derived from H$^{13}$CO$^+$ and $L'_{\rm mol}$). The solid lines are linear fits. The shadow regions represent 95\% confidence levels of the fits. (a) $L_{\rm bol}$ vs. $M_{\rm clump}$; (b) $L_{\rm bol}$ vs. $L'_{\rm HCN}$; (c) $L_{\rm bol}$ vs. $L'_{\rm HCO^+}$; (d) $L_{\rm bol}$ vs. $M_{\rm vir}$ derived from H$^{13}$CO$^+$; (e) $L_{\rm bol}$ vs. $L'_{\rm H^{13}CN}$; (c) $L_{\rm bol}$ vs. $L'_{\rm H^{13}CO^+}$.  \label{luminosity}}
\end{figure*}

% FIGURE 6 (as of April 22)
\begin{figure}
\centering
\includegraphics[width=\columnwidth]{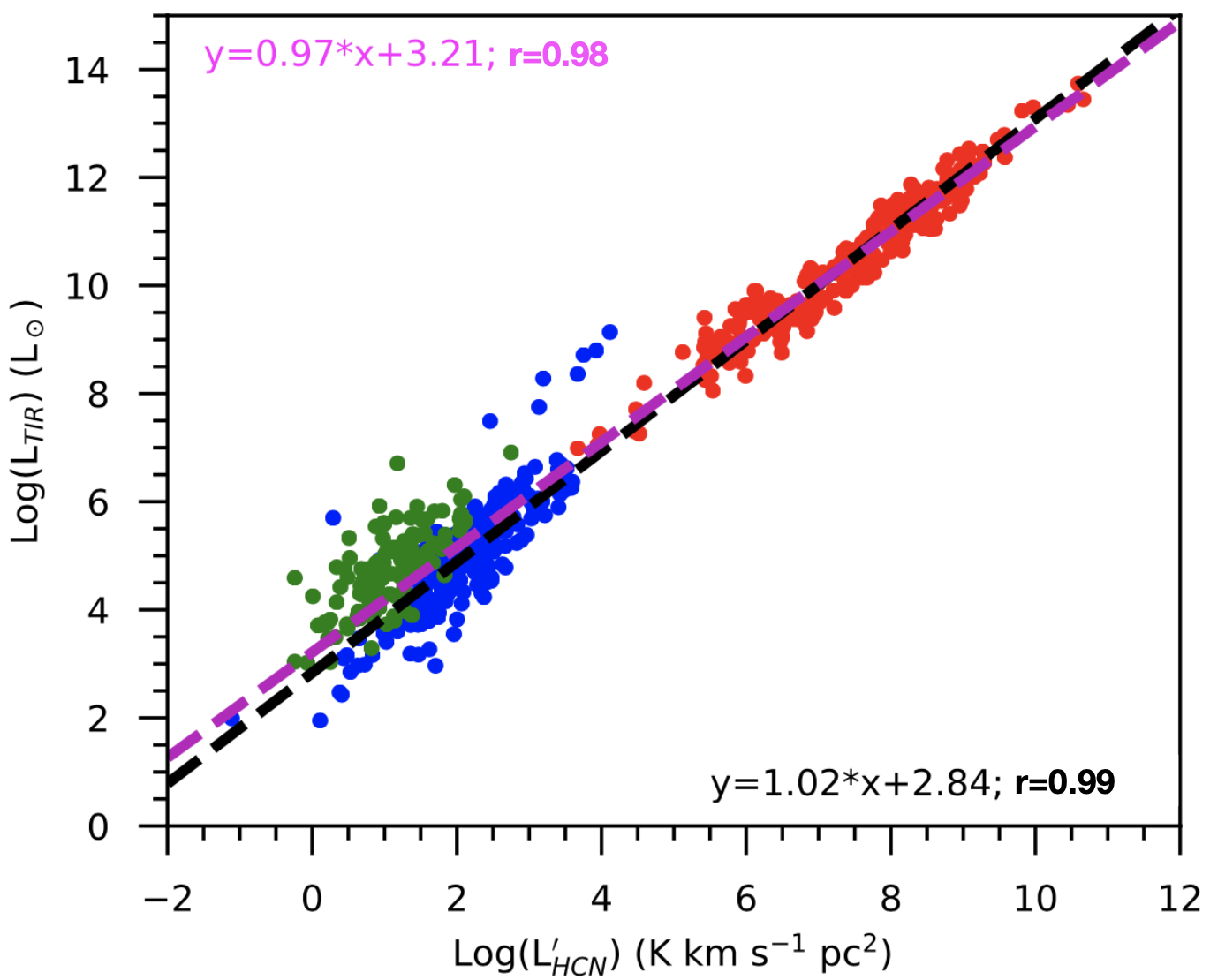}
\caption{The correlation between $L_{\rm TIR}$ and $L'_{\rm HCN}$. The data for compact objects in this study are shown as green filled circles. The data points for Galactic clumps (blue filled circles) and external galaxies (red filled circles) in single-dish observations were compiled by \citet{Jimenez2019}. The linear fit of all data is shown as the magenta dashed line. The linear fit toward single-dish data is shown as the black dashed line.  \label{extGAL}}
\end{figure}

In this section, we investigate how the infrared luminosities correlate with molecular line luminosities. \citet{Liu2016} has demonstrated that the bolometric luminosities ($L_{\rm bol}$) from SED fits are nearly identical to the total infrared luminosities ($L_{TIR}$) for this "ATOMS" sample. In this study, we still use $L_{\rm bol}$ \citep[from][]{Faundez2004,Urquhart2018} instead of $L_{\rm TIR}$. The molecular line luminosities ($L'_{\rm mol}$) are derived following \citet{Solomon1997} with the total integrated flux density. The $L_{\rm bol}$ and $L'_{\rm mol}$ are listed in Table \ref{LM}. The statistics like minumum, maximum, mean, median and standard deviation of the logarithmic values of $L_{\rm bol}$, $L'_{\rm mol}$ and their ratios are summarized in Table \ref{luminositypara}.

Figure \ref{luminosity} shows the correlations between $L_{\rm bol}$ and other parameters ($M_{\rm clump}$, $M_{\rm vir}$ derived from H$^{13}$CO$^+$, and $L'_{\rm mol}$). All the relationships appear approximately linear. The correlations are fitted with a linear function using the function "regplot" in the python package Seaborn\footnote{https://seaborn.pydata.org/generated/seaborn.regplot.html}. The slopes ($a$), intercepts ($b$) and correlation coefficients ($r$) from linear regressions are summarized in Table \ref{correlationpara}. Since the effective excitation densities of isotopologue lines are significantly larger than their main lines \citep{Shirley2015}, some models \citep[e.g.][]{Narayanan2008} would predict
that the correlations between $L_{\rm bol}$ and $L'_{\rm mol}$ are steeper for isotopologues than that for their main lines. In fact, the relations for isotopologues are not steeper and not even tighter for the "ATOMS" sample (see Figure \ref{luminosity}).

The mean log($L_{\rm bol}$/$L'_{\rm mol}$) ratios are 4.38, 4.49, 3.67, and 3.86 for  H$^{13}$CN, H$^{13}$CO$^+$, HCN and HCO$^+$, respectively. The $L_{\rm bol}$-to-$L'_{\rm HCN}$ and $L_{\rm bol}$-to-$L'_{\rm HCO^+}$ ratios for compact objects in the "ATOMS" sample seem to be a factor of $\sim$5 larger  than the corresponding ratios for Galactic clumps \citep[e.g.,][]{Wu2010,Stephens2016} and external galaxies \citep[e.g.,][]{Jimenez2019} measured with single-dishes. This difference is also understandable because the line emission in our ACA observations is mainly from compact and dense structures (likely to be more gravitationally bound objects). In contrast, line emission in other works with single-dishes also includes extended gas emission or even sub-thermal emission.

In Figure \ref{extGAL}, we plot the $L_{\rm TIR}$ (or $L_{\rm bol}$ ) as a function of the $L'_{\rm HCN}$ for the "ATOMS" measurements and literature measurements compiled by \citet{Jimenez2019} . The literature measurements include single-dish observations toward both Galactic clumps and external galaxies. This scaling relation between $L_{\rm TIR}$ and $L'_{\rm HCN}$ is nearly linear spanning almost 10 orders of magnitude in IR and HCN luminosity. The data points in the "ATOMS" measurements are clearly located above the relation determined by data points from single-dish measurements alone, indicating that the ACA observations in the "ATOMS" survey may reveal structures with higher star formation efficiency (SFE) in clouds. These structures in the ACA observations are much denser and more gravitationally bound than those revealed in single dish measurements (see Figure \ref{cmp}c).

\begin{table*}
	\centering
	\caption{Statistics of parameters of the 119 sources}
	\label{luminositypara}
	\begin{tabular}{cccccc} % four columns, alignment for each
		\hline
Parameters$^a$   &  Minimum  & Maximum  & Mean & Median & Std. Deviation\\
\hline
$R_{\rm eff}$                                &0.15    &4.26    &1.09     &0.84  &  0.75\\
$T_{\rm d}$                                  &19      &46      &29       &29    & 5 \\
$L_{\rm bol}$                                & 3.02   & 6.91   &	4.75  & 4.76&	0.77  \\
$M_{\rm clump}$                              & 1.59   & 5.04   &	3.24  & 3.19&	0.58  \\
$M_{\rm vir}^{13}$                           & 1.12   & 4.21   &	2.53  & 2.52&	0.50  \\
$M_{\rm vir}^{12}$                           & 1.86   & 4.86   &    3.30  & 3.31&   0.49 \\
$L'_{\rm H^{13}CN}$                          & -1.08  & 1.75   &	0.37  & 0.31&	0.57  \\
$L'_{\rm H^{13}CO^+}$                        & -1.34  & 1.87   &	0.16  & 0.14&	0.57  \\
$L'_{\rm HCN}$                               & -0.24  & 2.75   &	1.07  & 1.03&	0.57  \\
$L'_{\rm HCO^+}$                             & -0.62  & 2.87   &	0.89  & 0.89&	0.60  \\
$L_{\rm bol}$/$L'_{\rm H^{13}CN}$            & 3.43   & 6.63   &	4.38  & 4.32&	0.50  \\
$L_{\rm bol}$/$L'_{\rm H^{13}CO^+}$          & 3.26   & 6.53   &	4.59  & 4.57&	0.52  \\
$L_{\rm bol}$/$L'_{\rm HCN}$                 & 2.47   & 5.53   &	3.67  & 3.63&	0.53  \\
$L_{\rm bol}$/$L'_{\rm HCO^+}$               & 2.43   & 5.57   &	3.86  & 3.80&	0.53  \\
$L'_{\rm HCN}$/$L'_{\rm HCO^+}$              & -0.24  & 0.74   &	0.18  & 0.20&	0.18  \\
$L'_{\rm HCN}$/$L'_{\rm H^{13}CN}$           & -0.14  & 1.18   &	0.70  & 0.75&	0.27  \\
$L'_{\rm HCO^+}$/$L'_{\rm H^{13}CO^+}$       & -0.03  & 1.28   &	0.73  & 0.77&	0.26  \\
$L'_{\rm H^{13}CN}$/$L'_{\rm H^{13}CO^+}$    & -0.34  & 0.69   &	0.21  & 0.22&	0.20  \\
\hline
\multicolumn{6}{l}{$^a$ Except for $R_{\rm eff}$ and $T_{\rm d}$, the other parameters are logarithmic values. }\\
	\end{tabular}
\end{table*}

\begin{table}
	\centering
	\caption{Correlations between parameters}
	\label{correlationpara}
	\begin{tabular}{ccccccc} % four columns, alignment for each
		\hline
Relation   &  $a$  & $b$  & $r$  \\
\hline
$M_{\rm clump}$~--~$L_{\rm bol}$          &1.12(0.06)   & 1.12(0.21) & 0.83  \\
$M_{\rm vir}$~--~$L_{\rm bol}$            &1.04(0.11)   & 2.12(0.31) & 0.68  \\
$M_{\rm vir}$~--~$M_{\rm clump}$          &0.83(0.08)   & 1.14(0.22) & 0.73  \\
$L'_{\rm HCN}$~--~$L_{\rm bol}$           &0.99(0.08)   & 3.68(0.11) & 0.73  \\
$L'_{\rm HCO^+}$~--~$L_{\rm bol}$         &0.94(0.07)   & 3.91(0.09) & 0.73  \\
$L'_{\rm H^{13}CN}$~--~$L_{\rm bol}$      &1.03(0.08)   & 4.37(0.07) & 0.76  \\
$L'_{\rm H^{13}CO^+}$~--~$L_{\rm bol}$    &1.00(0.08)   & 4.59(0.05) & 0.74  \\
\hline
	\end{tabular}
\end{table}

\section{Discussion}

\subsection{The dispersion in the  $L_{\rm bol}$-to-$L'_{\rm mol}$ ratios}

\begin{figure}
\centering
\includegraphics[width=\columnwidth]{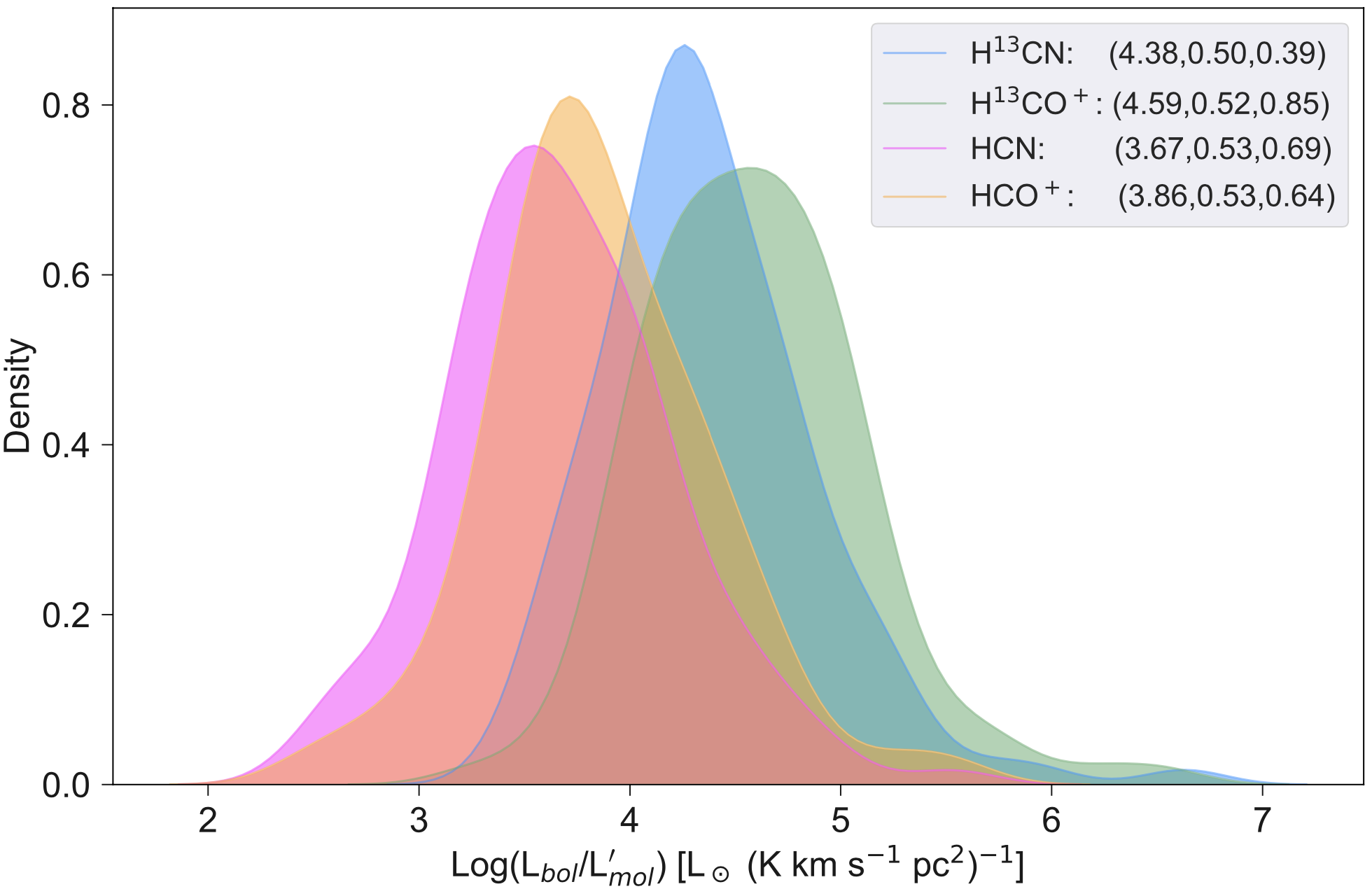}
\caption{Density distributions of  $L_{\rm bol}$-to-$L'_{\rm mol}$ ratios, plotted using the function "kdeplot" in the python package Seaborn. The numbers in brackets are mean, standard deviation and asymptotic significance in Kolmogorov-Sminov test of a normal distribution. \label{L-Ldist}}
\end{figure}

\begin{figure*}
\centering
\includegraphics[angle=0,scale=0.43]{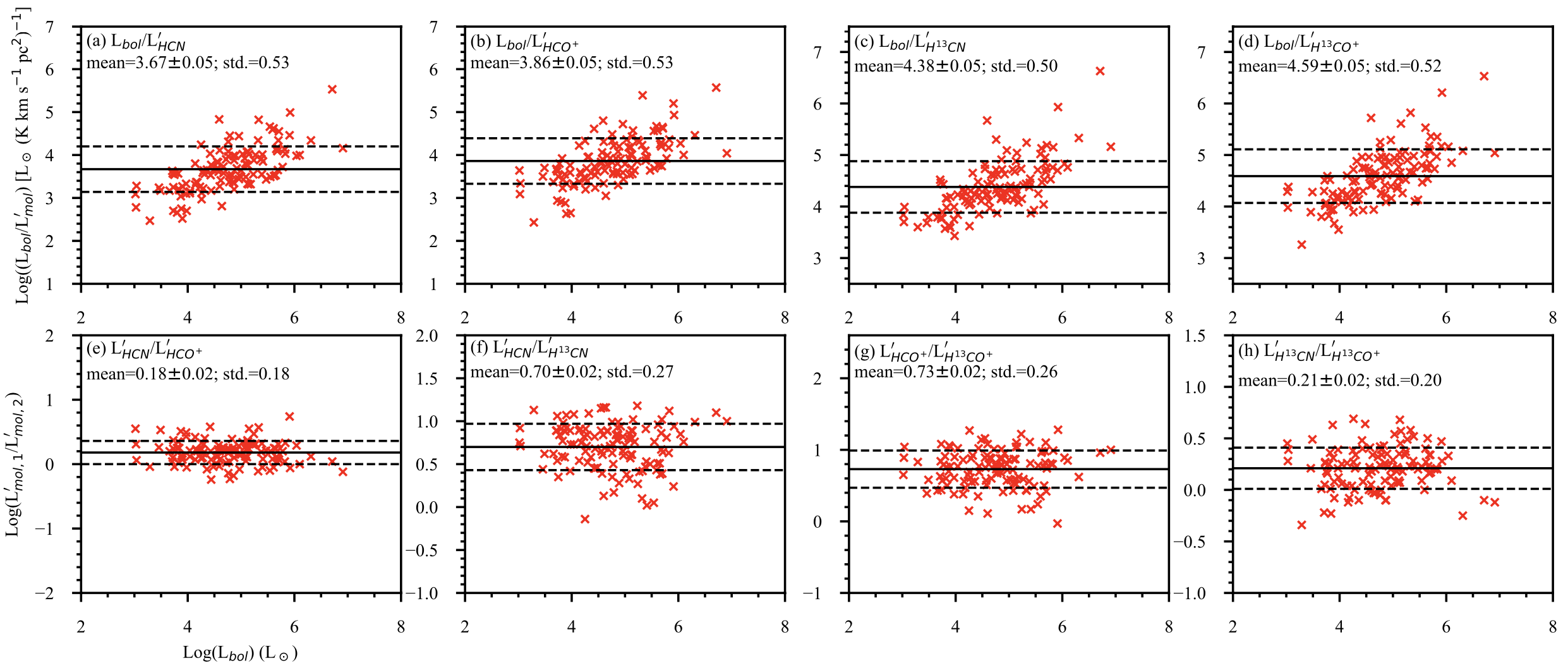}
\caption{ $L_{\rm bol}$-to-$L'_{\rm mol}$ ratios (panels a to d) and line luminosity ratios (panels e to h) as a function of bolometric luminosity $L_{\rm bol}$. The solid line is the mean value. The dashed lines show standard deviation. The name of the ratio is labeled at the upper-right corner in each panel. \label{lum-stat}}
\end{figure*}

\begin{figure*}
\centering
\includegraphics[angle=0,scale=0.43]{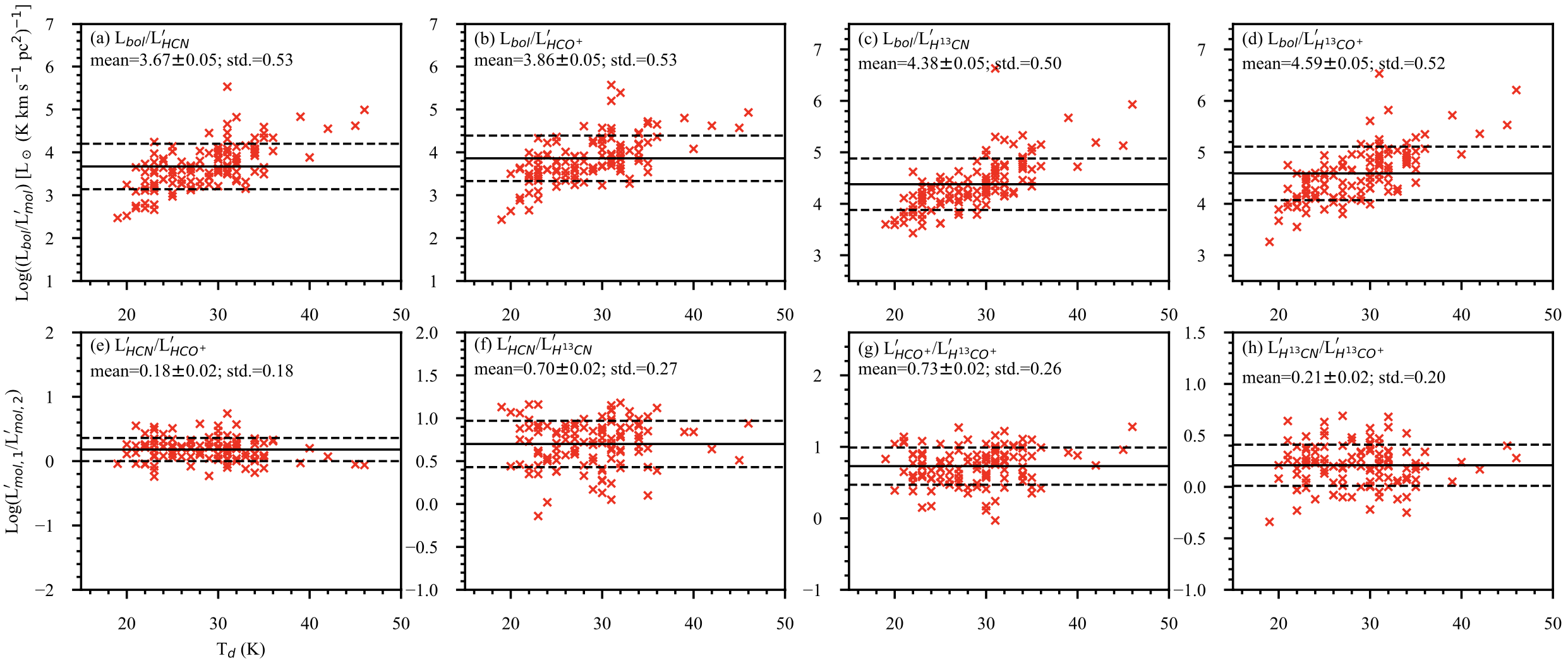}
\caption{ $L_{\rm bol}$-to-$L'_{\rm mol}$ ratios (panels a to d) and line luminosity ratios (panels e to h) as a function of dust temperature $T_{\rm d}$. The solid line is the mean value. The dashed lines show standard deviation. The name of the ratio is labeled at the upper-right corner in each panel. \label{Td-corr}}
\end{figure*}

\begin{figure*}
\centering
\includegraphics[angle=0,scale=0.43]{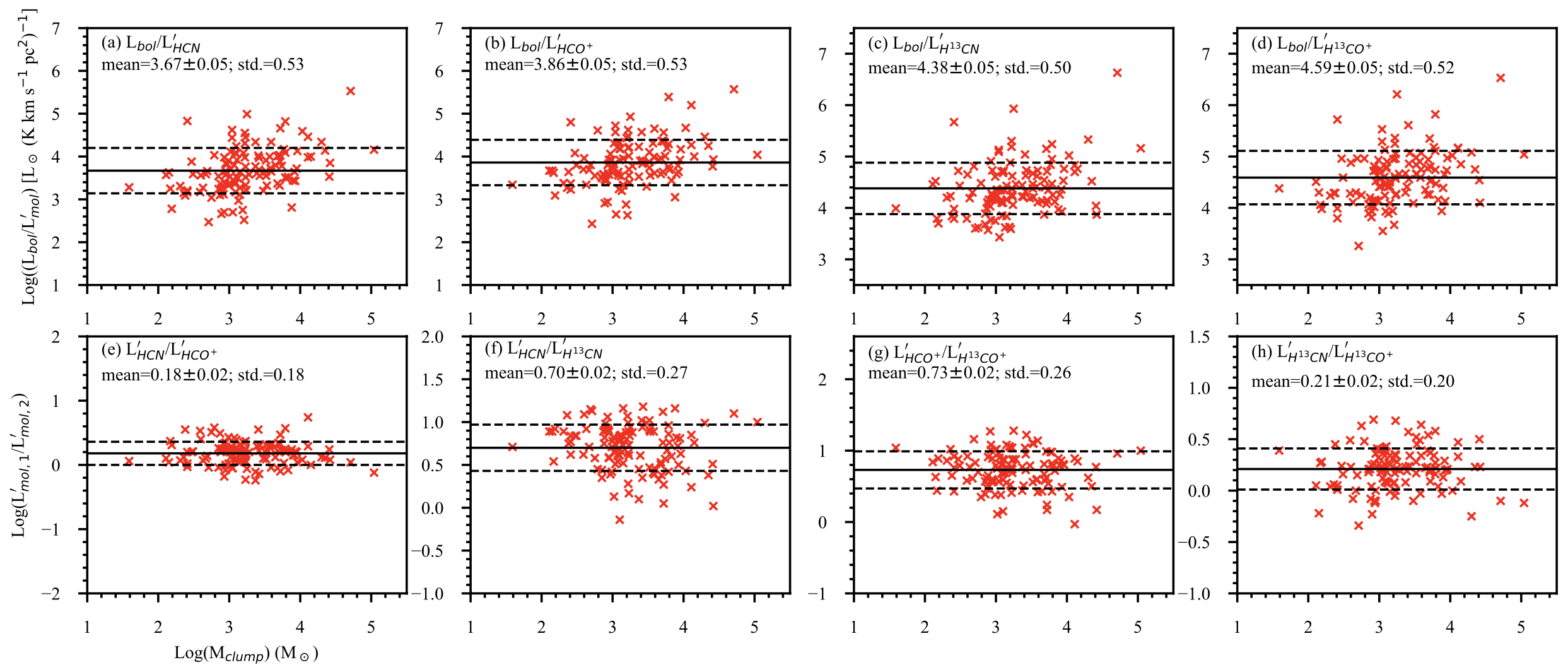}
\caption{ $L_{\rm bol}$-to-$L'_{\rm mol}$ ratios (panels a to d) and line luminosity ratios (panels e to h) as a function of clump masses $M_{\rm clump}$. The solid line is the mean value. The dashed lines show standard deviation. The name of the ratio is labeled at the upper-right corner in each panel. \label{mass-corr}}
\end{figure*}

\begin{figure*}
\centering
\includegraphics[angle=0,scale=0.43]{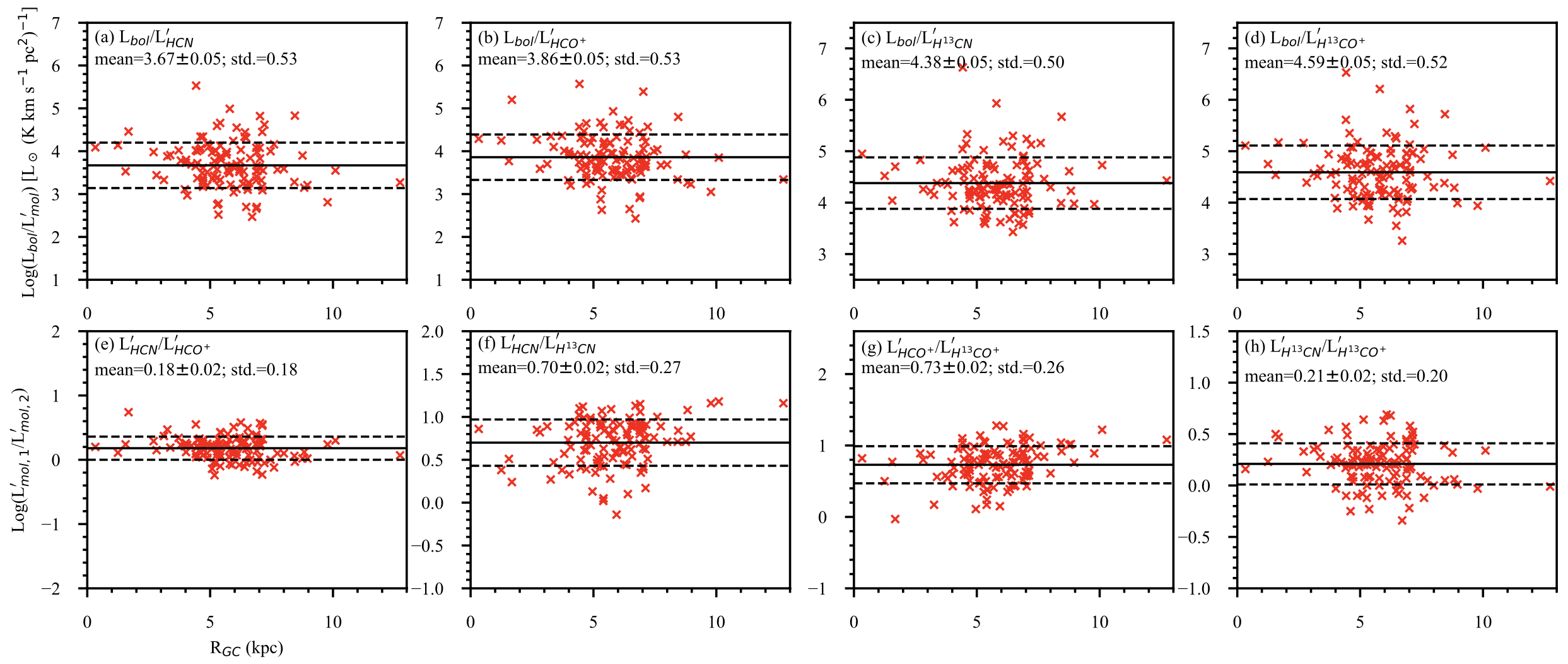}
\caption{$L_{\rm bol}$-to-$L'_{\rm mol}$ ratios and line luminosity ratios as a function of galactocentric distances. The name of the ratio is labeled at the upper-right corner in each panel. \label{RGC}}
\end{figure*}

\begin{figure}
\centering
\includegraphics[width=\columnwidth]{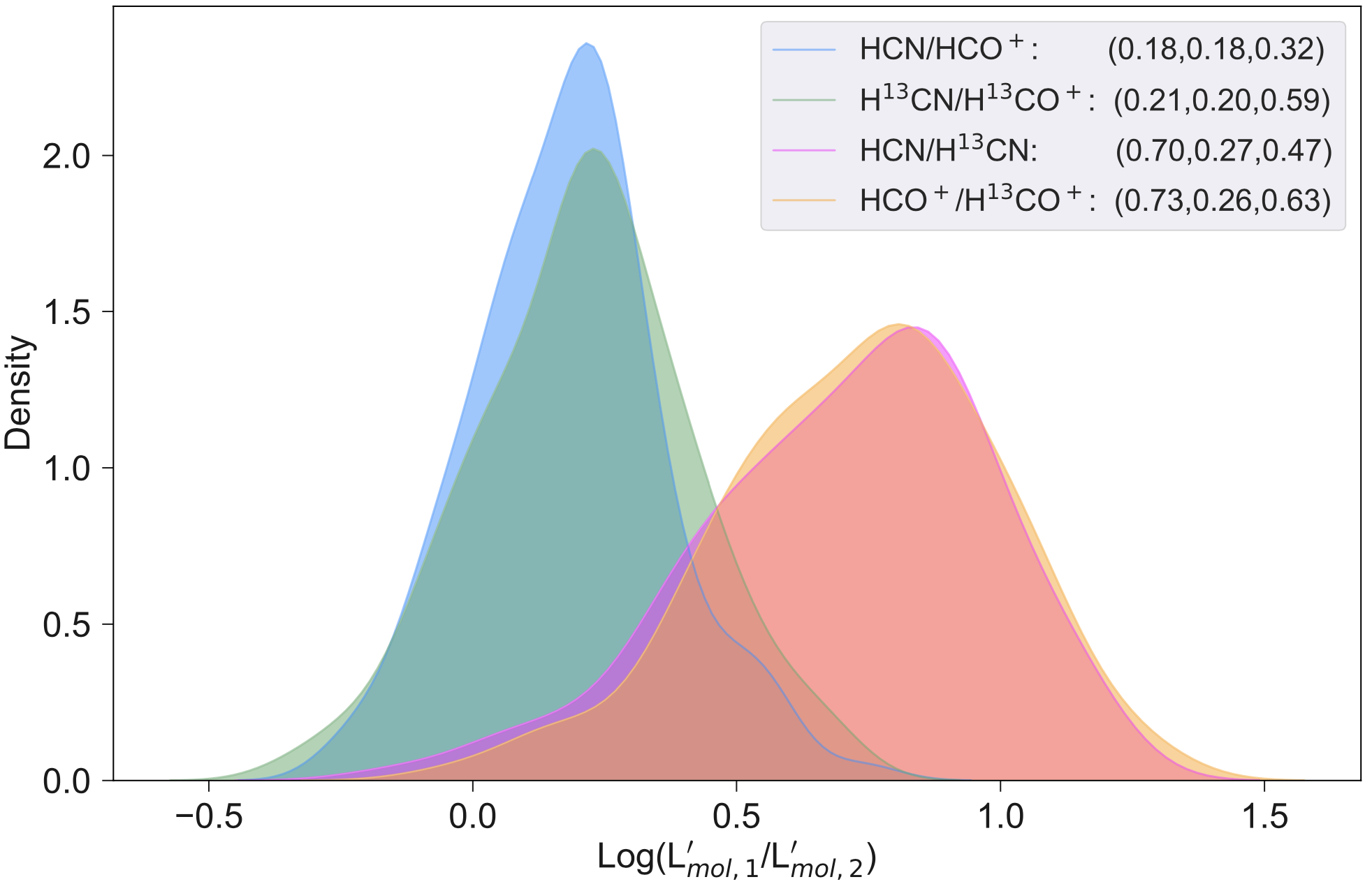}
\caption{Density distributions of  molecular line luminosity ratios, plotted using the function "kdeplot" in the python package Seaborn. The numbers in brackets are mean, standard deviation and asymptotic significance in Kolmogorov-Sminov test of a normal distribution. \label{linedist}}
\end{figure}

 We noticed a significant scatter in $L_{\rm bol}$-to-$L'_{\rm mol}$ ratios in our data as shown in Figure \ref{L-Ldist}. The standard deviations in distributions of log($L_{\rm bol}$/$L'_{\rm mol}$) are around 0.5 for all four molecular line transitions. The scatters in distributions for isotopologues are as large as for their main lines. We test whether the distributions of log($L_{\rm bol}$/$L'_{\rm mol}$) follow a normal distribution with Kolmogorov-Sminov test\footnote{https://www.spss-tutorials.com/spss-kolmogorov-smirnov-test-for-normality/}. The null hypothesis is that the distribution of log($L_{\rm bol}$/$L'_{\rm mol}$) follows a normal distribution. The tests retain the null hypothesis with asymptotic significances (or P-values; 2-tailed) $\ga$0.4 for the four lines (see Figure \ref{L-Ldist}), indicating that the distributions of log($L_{\rm bol}$/$L'_{\rm mol}$) likely follow a normal distribution. \citet{Stephens2016} also reported a significant scatter in $L_{\rm bol}$/$L'_{\rm HCN}$ at 1 pc clump-scale. They found that the difference between $L_{\rm bol}$/$L'_{\rm HCN}$ for the lowest 10\% quantile and the highest 90\% quantile is approximately two orders of magnitude. In our data, the difference is smaller, approximately one order of magnitude. It may imply that this ratio does not vary too much in gravitationally bound structures within clumps.

We have also investigated the similarities of distributions of log($L_{\rm bol}$/$L'_{\rm mol}$)  for different molecular line tracers. In Table \ref{test}, we test whether the Log($L_{\rm bol}$/$L'_{\rm mol}$) values for two molecular lines could be drawn from the same distribution with Kolmogorov-Sminov test after shifting the distributions by their mean values. The null hypothesis in tests is the shapes of the distributions for Log($L_{\rm bol}$/$L'_{\rm mol}$) values are the same for the two molecular lines. With non-parametric tests, we found that the distributions of log($L_{\rm bol}$/$L'_{\rm mol}$)  for different molecular line tracers are very similar. In  particular, the distributions of Log($L_{\rm bol}$/$L'_{HCN}$) and Log($L_{\rm bol}$/$L'_{HCO^+}$) show the highest similarity with asymptotic significances $\sim$0.97 in Kolmogorov-Sminov test. The distributions for isotopologues (H$^{13}$CN and H$^{13}$CO$^+$) are also similar to the distributions for their corresponding main lines (HCN and HCO$^+$) with asymptotic significance as high as $\sim$0.95 in Kolmogorov-Sminov test. This implies that although the main lines may not trace well the dense structures (cores or filaments) within clumps/clouds \citep[e.g.,][]{Kauffmann2017,Pety2017,Shimajiri2017,Liu2020}, they can still reveal the total dense gas masses as well as their isotopologues in statistics.

Panels (a)  to (d) in Figure \ref{lum-stat} present correlations between the $L_{\rm bol}$-to-$L'_{\rm mol}$ ratios and $L_{\rm bol}$. \citet{Wu2005,Wu2010} suggest that above an infrared luminosity threshold of $\sim10^{4.5}$ $L_{\sun}$, the $L_{\rm bol}$-to-$L'_{\rm mol}$ ratios become constant. However, we do not see such threshold in our data. There seems to be a clear trend that $L_{\rm bol}$-to-$L'_{\rm mol}$ ratio increases with $L_{\rm bol}$, spanning four orders of magnitude in $L_{\rm bol}$. This is likely caused by evolutionary effects of the sources, as suggested by \citet{Liu2016} and \citet{Stephens2016}. \citet{Liu2016} reported a bimodal behavior in the $L_{\rm bol}$-to-$L'_{\rm mol}$ correlations for clumps with different dust temperature, luminosity-to-mass ratio, and virial parameter. More luminous (or more evolved) sources seem to have consumed more gas, leading to higher $L_{\rm bol}$-to-$L'_{\rm mol}$ ratios.

More evolved star forming regions might have a higher dust temperature. To test the evolutionary effect, we present in Figure \ref{Td-corr} correlations between the $L_{\rm bol}$-to-$L'_{\rm mol}$ ratios and dust temperature $T_{\rm d}$. As expected, there is a clear trend that $L_{\rm bol}$-to-$L'_{\rm mol}$ increases with T$_d$. The trend is even more clearly seen for isotopologues. Same trend is seen in Figure \ref{lum-stat}(a-d) as well where the trend is also more clearly seen in isotopologues.

Clump masses are not as sensitive to the evolutionary effect as bolometric luminosity or dust temperature. Panels (a)  to (d) in Figure \ref{mass-corr} present correlations between the $L_{\rm bol}$-to-$L'_{\rm mol}$ ratios and $M_{\rm clump}$. Indeed, there is no clear trend between $L_{\rm bol}$-to-$L'_{\rm mol}$ ratios and $M_{\rm clump}$ that spans three orders of magnitude in $M_{\rm clump}$.

Panels (a)  to (d) in Figure \ref{RGC} present correlations between the $L_{\rm bol}$-to-$L'_{\rm mol}$ ratios and galactocentric distances ($R_{\rm GC}$) for different tracers. \citet{Jimenez2019} found that $L_{\rm TIR}$-to-$L'_{\rm HCN}$ tends to increase with increasing $R_{\rm GC}$ in most of their targeted external galaxies, but with large galaxy-to-galaxy scatter. In our data, however, there is no clear trend for $L_{\rm bol}$-to-$L'_{\rm mol}$ ratios against $R_{\rm GC}$, suggesting that this ratio may be constant for gravitationally bound clumps in different Galactic environments.

\begin{table}
	\centering
	\caption{Asymptotic significances (p-value) in non-parametric tests}
	\label{test}
	\begin{tabular}{lc} % four columns, alignment for each
		\hline
		Distributions    & Kolmogorov-Sminov$^a$ \\
		\hline
log($\frac{L_{bol}}{\rm H^{13}CN}$) vs. log($\frac{L_{bol}}{\rm H^{13}CO^+}$)      & 0.69 \\
log($\frac{L_{bol}}{\rm HCN}$) vs. log($\frac{L_{bol}}{\rm HCO^+}$)                &0.97   \\
log($\frac{L_{bol}}{\rm H^{13}CN}$) vs. log($\frac{L_{bol}}{\rm HCN}$)		         &0.95 \\
log($\frac{L_{bol}}{\rm H^{13}CO^+}$) vs. log($\frac{L_{bol}}{\rm HCO^+}$)          & 0.95  \\
log($\frac{\rm HCN}{\rm HCO^+}$) vs. log($\frac{\rm H^{13}CN}{\rm H^{13}CO^+}$)            & 0.89  \\
log($\frac{\rm HCN}{\rm H^{13}CN}$) vs. log($\frac{\rm HCO^+}{\rm H^{13}CO^+}$)            & 0.89 \\
		\hline
\multicolumn{2}{l}{$^a$ The significance level ($\alpha$-value) is 0.05. }\\
	\end{tabular}
\end{table}

\subsection{Variations of molecular line luminosity ratios}

Figure \ref{linedist} plots the density distributions of various molecular line luminosity ratios (or integrated intensity ratios). Interestingly, we found that the distribution of HCN-to-HCO$^+$ ratios is very similar to the distribution of H$^{13}$CN-to-H$^{13}$CO$^+$ ratios. The non-parametric tests in Table \ref{test} indicate that they can be drawn from a same distribution. It implies that the optical depths of the main lines do not affect the determination of HCN-to-HCO$^+$ ratios statistically. The mean HCN-to-HCO$^+$ ratios inferred from distributions of log($\frac{\rm HCN}{\rm HCO^+}$) and log($\frac{\rm H^{13}CN}{\rm H^{13}CO^+}$) are $\sim$1.5 and $\sim$1.6, respectively. This mean HCN-to-HCO$^+$ ratio (1.5-1.6) is slightly larger than that in nearby disk galaxies \citep[1.3-1.4;][]{Jimenez2019}, Orion B GMC \citep[0.9-1.1;][]{Pety2017,Shimajiri2017} and Galactic Plane GMCs \citep[1-1.4;][]{Nguyen-Luong2020,Wang2020}. The mean HCN-to-HCO$^+$ ratio in the ``ATOMS" sample is similar to that of Aquila GMC \citep[1.6;][]{Shimajiri2017}. The line ratio of HCN-to-HCO$^+$ seems to be sensitive to environments \citep{Pety2017,Shimajiri2017}. High HCN-to-HCO$^+$ ratios have been found in far-UV irradiated environments such as evolved Galactic H{\sc ii} regions \citep{Nguyen-Luong2020}, AGNs \citep{Aladro2015} or luminous infrared galaxies \citep[LIRGs;][]{Papadopoulos2007}. This implies that HCO$^+$ abundance is sensitive to the ionization degree of molecular gas. In more far-UV irradiated environments, HCO$^+$ is more easily recombined with free electrons, leading to a decrease in its abundance. Increasing HCN-to-HCO$^+$ ratios with increasing infrared luminosity have been found in LIRGs \citep{Papadopoulos2007}. In the Galaxy, more evolved H{\sc ii} regions also show higher HCN-to-HCO$^+$ ratios than infrared dark clouds \citep{Nguyen-Luong2020}. However, in our data, we do not see any trend of HCN-to-HCO$^+$ ratios against the $L_{\rm bol}$ spanning four orders of magnitude. This may imply that the majority of HCN and HCO$^+$ J=1-0 line emission within the "ATOMS" clumps comes from regions that are not exposed to strong local FUV radiation field. This scenario needs to be tested from detailed analysis of the spatial distributions of molecular line emission inside the clumps, which will be presented in forthcoming works using the high resolution ($\sim2\arcsec$) ALMA 12-m array data of the "ATOMS" survey \citep[see][for example]{Liu2020}.

The density distributions for log($\frac{\rm HCN}{\rm H^{13}CN}$) and log($\frac{\rm HCO^+}{\rm H^{13}CO^+}$) are also quite similar to each other but with large scatter ($\sigma\sim$0.6) in the data (see Figure \ref{linedist}). The large scatter means that the optical depths of the main lines show significant variations among sources. The non-parametric tests in Table \ref{test} confirm that their distributions are nearly the same after shifting the distributions by their mean values, indicating that HCN $J=1-0$ and HCO$^+$ $J=1-0$ may have similar opacity in sources of the ``ATOMS" sample. The mean HCN-to-H$^{13}$CN ratio and HCO$^+$-to-H$^{13}$CO$^+$ ratio are $\sim$5.0 and $\sim$5.4, respectively. Assuming an isotopic $^{12}$C/$^{13}$C ratio of 55 at the median galactocentric distance of 5.8 kpc \citep{Milam2005}, the representative optical depths inferred from these mean ratios are $\sim$12 and $\sim$11 for HCN $J=1-0$ and HCO$^+$ $J=1-0$, respectively. Thus the column densities of HCN and HCO$^+$ derived from the main lines with optically thin assumption could be underestimated by a factor of $\sim$10 for such high optical depths.

However, there is no trend between molecular line luminosity ratios (HCN-to-HCO$^+$, H$^{13}$CN-to-H$^{13}$CO$^+$, HCN-to-H$^{13}$CN ratio and HCO$^+$-to-H$^{13}$CO$^+$) and $L_{\rm bol}$ (see panels (e) to (h) in Figure \ref{lum-stat}). All these molecular line luminosity ratios seem constant against $L_{\rm bol}$ spanning 4 orders of magnitude in $L_{\rm bol}$.  This implies that optical depths (or abundance ratios) of main lines should not affect the interpretation of the slopes in star formation relations. The molecular line luminosity ratios are also not correlated with dust temperature (see panels (e) to (h) in Figure \ref{Td-corr}) or clump masses (see panels (e) to (h) in Figure \ref{mass-corr}).

Panels (e)  to (h) in Figure \ref{RGC} present correlations between molecular line luminosity ratios and galactocentric distances ($R_{\rm GC}$) for different tracers. In a survey of nine nearby external galaxies, \cite{Jimenez2019} found that there is a decreasing trend of HCN-to-HCO$^+$ as a function of galactocentric distances in only two galaxies. No significant changing HCN-to-HCO$^+$ ratios are witnessed in the other seven galaxies. There is also no obvious trend for molecular line luminosity ratios against $R_{\rm GC}$ in our data, indicating similar local environments (such as FUV radiation field, density, temperature) for the sources in the "ATOMS" sample.

\subsection{Star Formation Efficiencies}\label{sfe}

\begin{table*}
\centering
\caption{Statistics of the log of Star Formation Efficiency} \label{tabsfe}
\begin{tabular}{c c c c c c c c c ccccc l }
\hline
\hline
 && \multicolumn{3}{c}{SFR/M$_{\rm vir}$(HCO$^+$)} & &\multicolumn{3}{c}{SFR/M$_{\rm clump}$} & &\multicolumn{3}{c}{SFR/M$_{\rm vir}$(H$^{13}$CO$^+$)} & \cr
\cline{3-5}\cline{7-9}\cline{11-13}
Number && Mean & Median & Std &&  Mean & Median & Std &&  Mean & Median & Std && Selection$^a$ \cr
\hline
119 && $-2.38$ & $-2.33$ & $0.63$ && $-2.33$ & $-2.29$ & $0.43$ &&  $-1.62$  & $-1.66$ & $0.55$ &&   All \cr
75 && $-2.09$ & $-2.11$ & $0.50$ && $-2.13$ & $-2.10$ & $0.36$ &&  $-1.35$ &  $-1.39$ & $0.46$ &&   SFR$>5$ \cr
75 && $-2.09$ & $-2.11$ & $0.50$ && $-2.13$ & $-2.10$ & $0.36$ &&  $-1.35$ &  $-1.39$ & $0.46$ &&   SFR$>5$, $r > 0.2$ \cr
44 && $-2.89$ & $-2.79$ & $0.50$ && $-2.66$ & $-2.69$ & $0.32$ &&  $-2.08$ &  $-2.13$ & $0.35$ &&   SFR$<5$, $r > 0.2$ \cr
\hline
\multicolumn{4}{l}{$^a$ Selection criteria explained in the text. }\\
\end{tabular}
\end{table*}

Until now, we have dealt with observables, comparing luminosities and masses of various
quantities. Now, we engage in the more speculative issues of star formation rate (SFR) and star formation efficiency (SFE). We use the extragalactic definition of SFE: the SFR per unit mass of gas, with units of Myr$^{-1}$. On the scale of galaxies and averaged over 5 Myr or longer, the far-infrared luminosity can measure the star formation rate. The relation given in Table 1 of
\citet{Kennicutt2012},
and explained in more detail in
%Hao et al. 2011, Murphy et al. 2011
\citet{2011ApJ...741..124H} and
\citet{2011ApJ...737...67M},
can be written in convenient units as

\begin{equation}
SFR = 1.49 \times 10^{-4} L_{\rm bol}(L_{\sun})\  \msun\ {\rm Myr}^{-1}
\end{equation}

This conversion depends on assumptions about IMF and star formation history, so there is no guarantee that it would apply to the spatial and temporal scales studied in this paper. However, we wish to compare the current results to those of other works that used different indicators of star formation rate. Consequently, we use the equation above to generate estimates of SFR and compare to various mass estimates to gauge the SFE. Comparison of SFE, as opposed to SFR, among samples is best suited to appreciate the dispersion and the relative quality of different predictors of star formation, as shown by \citet{Vutisalchavakul2016}; so we focus on SFE here. We further follow \citet{Vutisalchavakul2016} in computing means, medians, and standard deviations of SFE in the log, and present them in Table \ref{tabsfe}.

As predictors, we consider three estimates of the dense gas mass. These estimates also come with various caveats about what characteristic density they represent
\citep{Evans2020}. The first estimate is the masses ($M_{\rm clump}$ in Table \ref{LM}) of clumps from single dish observations, which were obtained from fitting SEDs using far infrared (IRAS or Herschel) data and (sub-)millimeter continuum data (0.87 mm from \citet{Urquhart2018} or 1.2 mm from \citet{Faundez2004}). We also use, secondly, the virial masses derived from both HCO$^+$ and H$^{13}$CO$^+$ ($M_{\rm vir}^{12}$ and $M_{\rm vir}^{13}$in Table \ref{LM}).

We compare our results for SFE to those of \citet{Vutisalchavakul2016} first. They determined the SFR from the mid-infrared luminosity, again using the extra-galactic relations, which carry the same caveat about IMF and star formation history discussed above. However, they did find that the SFR from the mid-infrared luminosity agreed well with those from the free-free radio continuum, which averages over times of 3 to 10 Myr, as long as the inferred SFR was at least 5 \msun\ Myr$^{-1}$. The criterion of log $L_{\rm bol} \geq 4.5$ used by  \citet{Wu2010} implies a star formation rate of $4.7$ \msun\ Myr$^{-1}$, essentially the same criterion. Of the 119 sources in Table \ref{LM}, 75 satisfy this criterion. Also the sample of \citet{Vutisalchavakul2016} was defined by submillimeter continuum emission, which traced relatively large, somewhat dense clumps. If we make the cut between cores and clumps at a size of 0.2 pc (based on $R_{\rm eff}$ in Table 6), all 75 sources that made the first cut also satisfy the size criteria that make them directly comparable to the sample of
\citet{Vutisalchavakul2016}.

\begin{figure}
\centering
\includegraphics[width=\columnwidth]{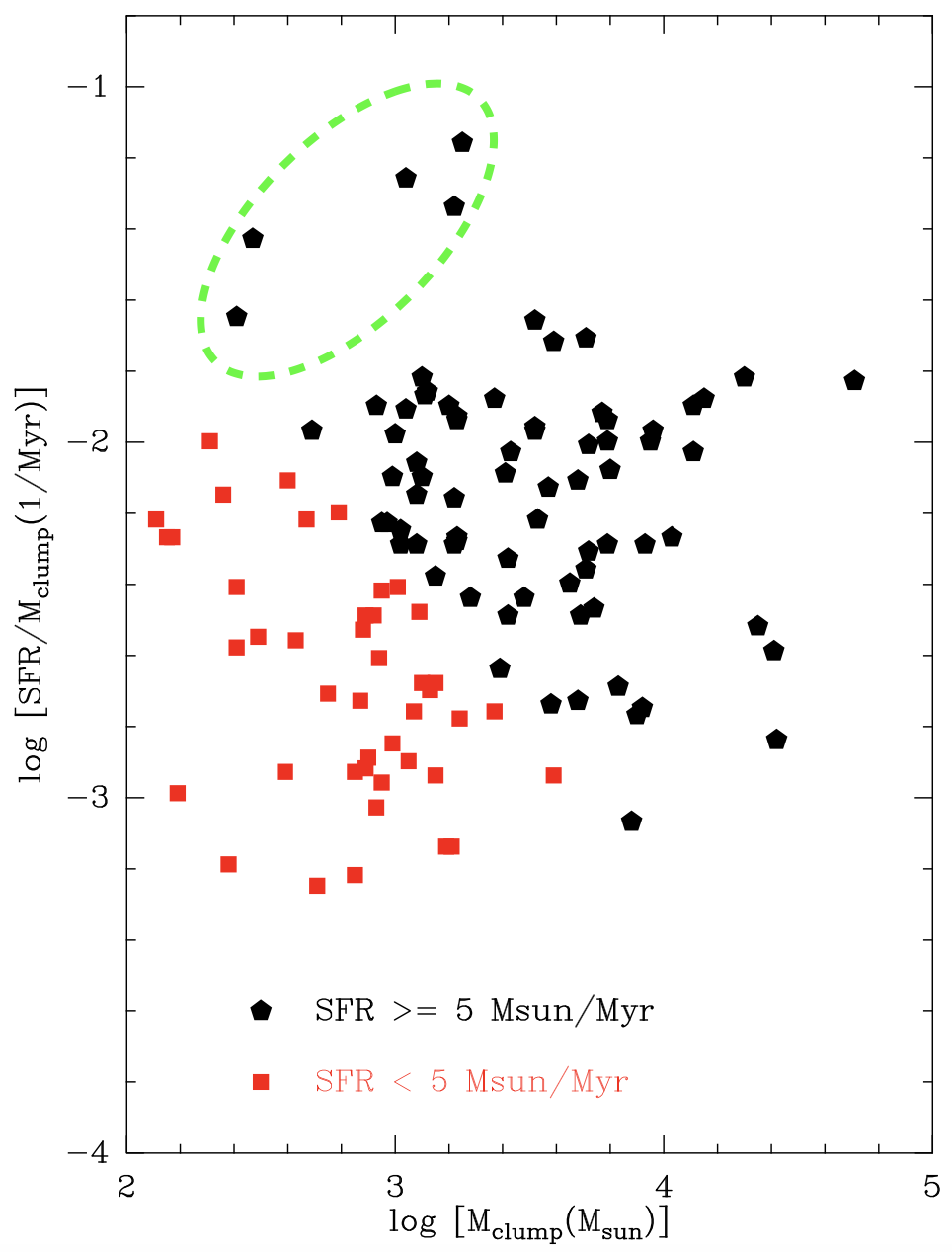}
\caption{SFR/$M_{\rm clump}$ is plotted versus $M_{\rm clump}$. The red points represent the sources with SFR $< 5$ \msun\ Myr$^{-1}$, while the black points represent the sources with  SFR $\geq 5$ \msun\ Myr$^{-1}$. The points inside the green dashed ellipse are five outliers with highest dust temperature in the sample.
\label{sfevmcl}
}
\end{figure}

Figure \ref{sfevmcl} plots the log of SFE versus the log of $M_{\rm clump}$ for sources that do (black) and do not (red) meet the requirement on SFR of  5 \msun\ Myr$^{-1}$.
There are no obvious overall trends in SFE within each sample, but the low luminosity sources lie clearly below the ones above the threshold previously used  \citep{Wu2010, Vutisalchavakul2016}. There are five clear outliers (I09002-4732, I12320-6122, I16562-3959, I17258-3637 and I18139-1842) in Figure \ref{sfevmcl} at high SFE. These turn out to have unusually large values for $T_d$ ($\ga$39 K), as listed in Table \ref{LM}. To the extent that $T_d$ reflects evolutionary state, these sources would be more evolved. If so, a higher fraction of their masses would have been converted into stars, leading to a higher SFE. Table \ref{tabsfe} shows the logarithmic means, medians, and standard deviations  of the full sample, along with those with the restrictions applied, for three estimates of the mass of dense gas. While the cuts decrease the means and medians, the differences are not large (about 0.3 dex). The last line in Table \ref{tabsfe} shows the statistics for the 44 sources that lie below the cut in star formation rate of 5 \msun\ Myr$^{-1}$; their mean and median SFE are much lower, reflecting the fact that log $L_{\rm bol}$ no longer traces the SFR below that value.  The SFE based on the  HCO$^+$ virial mass and the single-dish clump mass are quite similar, but the latter has a smaller dispersion (0.36 dex). The SFE based on the virial mass from H$^{13}$CO$^+$ are larger by about 0.8 dex. The mean value for the log of SFE from \citet{Vutisalchavakul2016} is $-1.74 \pm 0.50$ for the dense gas, with mean volume density of $n \sim 10^3$ cm$^{-3}$, 0.38 dex higher than the current sample. In turn, the sample of nearby clouds, where the dense gas mass was determined from the condition that $A_V \geq 8$ mag, had a still higher SFE, about $-1.61$. The SFR in the nearby clouds was measured by YSO counting, which is much less sensitive to the IMF and star formation history
\citep{2014ApJ...782..114E}.

\begin{figure}
\centering
\includegraphics[width=\columnwidth]{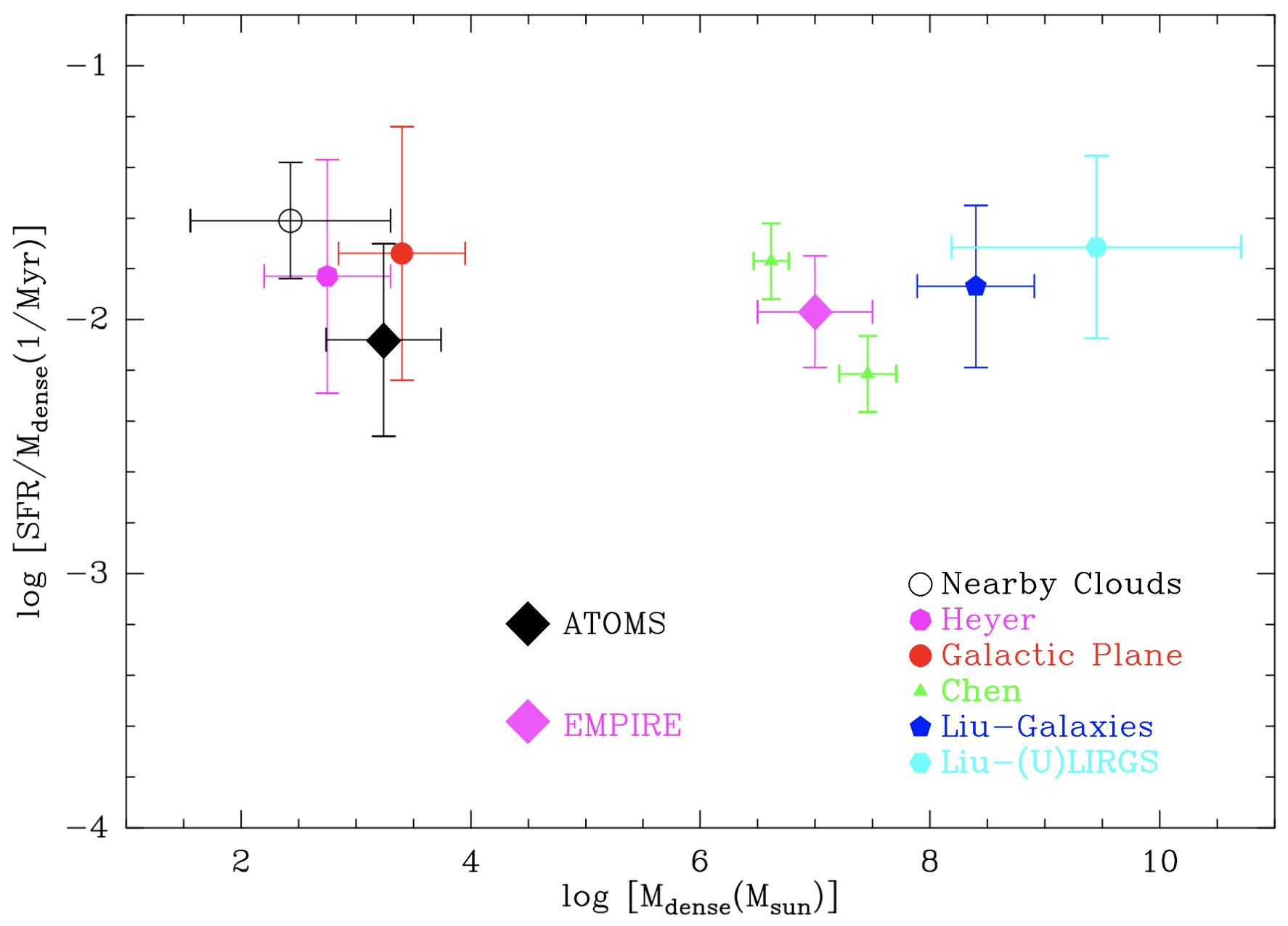}
\caption{The logarithmic mean SFE (SFR/$M_{\rm dense}$) is plotted versus $M_{\rm dense}$ with both showing the standard deviation in the log. The black diamond is from this work, and the magenta diamond is from the EMPIRE study of nearby galaxies \citep{Jimenez2019}. The other points are the same as those in \citet{Vutisalchavakul2016}.The red filled circle is from \citet{Vutisalchavakul2016} for Galactic Plane clouds, the empty black circle is from \citet{2014ApJ...782..114E} for nearby clouds. The magenta
point (heptagon) comes from \citet{Heyer2016} for a sample of molecular clouds containing 24 $\micron$-based Class I protostars. The two green triangles are from \citet{Chen2015} for resolve giant molecular clouds in M51 (lower one for inner galaxy and higher one for outer galaxy). The blue and cyan points represent the
whole normal galaxies and the whole (U)LIRGs, respectively, from \citet{Chen2015}.
\label{galmdense}
}
\end{figure}

The masses of dense gas for this sample ($\mean{\log(M_{\rm clump})} = 3.24$; see Table \ref{luminositypara}) seem to be similar to that of \citet{Vutisalchavakul2016} ($\mean{\log(M_{\rm clump})} \sim 3.4$). However the sizes of the regions are somewhat smaller with a mean effective clump radius of $1.1$ pc (see Table \ref{luminositypara}), versus $3.1 \pm 2.1$ pc for the sample of \citet{Vutisalchavakul2016} . Consequently the surface  and volume densities are higher. The sample selection may play a role. This sample was selected from a survey of CS $J = 2-1$ emission (though that survey was itself directed towards IRAS point sources with infrared colors similar to UC H{\sc ii} regions), thus focusing on relatively dense gas, while the sample of \citet{Vutisalchavakul2016} began with surveys of radio recombination lines, requiring recent formation of massive stars. Based on smaller sizes and higher surface densities, this sample may be younger than the sample of \citet{Vutisalchavakul2016}
so a smaller SFE would be expected.

Despite the differences, the values from this paper lie well within the dispersion of values from \citet{Vutisalchavakul2016} and are consistent with the nearly constant SFE from nearby clouds to distant galaxies shown in figure 12 of \citet{Vutisalchavakul2016}. We reproduce that figure here (see Figure \ref{galmdense}), with the points for ATOMS and for the EMPIRE study \citep{Jimenez2019} added.

\section{Conclusions}

Some molecular gas tracers (e.g., HCN and HCO$^{+}$) that are commonly used in studies of external galaxies have large optical depths and may not be good tracers of the distribution of dense molecular gas in molecular clouds. In this work, we use the  isotopologues of HCN and HCO$^{+}$ to calibrate the relationships between the recent star formation rate (SFR), as traced by the total infrared emission, and the dense molecular gas mass determined from line luminosities ($L'_{\rm mol}$). The data used in this work are from ACA observations in the ``ATOMS" survey of 146 active Galactic star forming regions. Our main results are summarized as follows:

(1) We extracted 173, 184, 190, 189 and 182 compact objects from 3 mm continuum emission, H$^{13}$CN $J=1-0$, H$^{13}$CO$^+$ $J=1-0$, HCN $J=1-0$, and HCO$^+$ $J=1-0$ line emission, respectively. The compact objects in 3 mm continuum emission are systematically smaller than compact objects in molecular line emission. Compact sources in HCN $J=1-0$, and HCO$^+$ $J=1-0$ line emission have statistically larger effective radii than the compact objects found in the line emission of their isotopologues.

(2)  The virial masses of compact objects in H$^{13}$CO$^+$ $J=1-0$ line emission are systematically smaller than the total masses of their clumps with a median virial parameter ($\alpha$=$M_{\rm vir}$/$M_{\rm clump}$) of $\sim$0.2, While the virial masses from HCO$^+$ are comparable to the total masses of their natal clumps with a median virial parameter of $\sim$1.2. It suggests that our ACA observations most likely detect the gravitationally bound structures inside clumps.

(3) All the correlations between $L_{\rm bol}$, tracing the SFR, and molecular line luminosities $L'_{\rm mol}$ of the four transitions (H$^{13}$CN $J=1-0$, H$^{13}$CO$^+$ $J=1-0$, HCN $J=1-0$, and HCO$^+$ $J=1-0$)  appear approximately linear. Line emission of isotopologues shows scatter in their $L_{\rm bol}$-$L'_{\rm mol}$ relations as large as that found in their main line emission. Although the main lines may not be good tracers of the spatial distribution of dense gas in molecular clouds \citep{Pety2017,Shimajiri2017}, they trace the total dense gas masses as well as do their isotopologues.

(4) The density distributions of log($L_{\rm bol}$/$L'_{\rm mol}$)  for different molecular line tracers show high similarity, indicating that they can be drawn from a same underlying distribution, again confirming that the main lines trace the total dense gas masses as well as do their isotopologues. There seems to be a clear trend between increasing $L_{\rm bol}$-to-$L'_{\rm mol}$ ratios and $L_{\rm bol}$ (also $T_{\rm d}$) spanning four orders of magnitude in $L_{\rm bol}$. This is likely caused by evolutionary effects in the sources. $L_{\rm bol}$-to-$L'_{\rm mol}$ ratios do not vary with galactocentric distance ($R_{\rm GC}$), suggesting that SFEs remain constant for gravitationally bound clumps in different Galactic environments.

(5)  All the molecular line luminosity ratios (HCN-to-HCO$^+$, H$^{13}$CN-to-H$^{13}$CO$^+$, HCN-to-H$^{13}$CN ratio and HCO$^+$-to-H$^{13}$CO$^+$) appear constant when compared to $L_{\rm bol}$ over four orders of magnitude in $L_{\rm bol}$.  This implies that the large optical depths (or abundance ratios) of main lines do not affect the interpretation of the slopes in star formation relations. There is also no obvious overall trend for molecular line luminosity ratios against $T_{\rm d}$, $M_{\rm clump}$, and $R_{\rm GC}$ in our data.

(6) If we use extragalactic calibrations to convert  $L_{\rm bol}$ to a star formation rate and compare it to masses of clumps from single-dish observations or to  virial masses from HCO$^+$, we can estimate a dense gas star formation efficiency. While this procedure is not fully justified for the scales we are studying, we find that the logarithmic mean SFE is reasonably consistent with other measures of SFE for dense gas, even those using very different tracers or examining very different spatial scales.

\section*{Acknowledgements}

Tie Liu is supported by the initial funding of scientific research for high-level talents at Shanghai Astronomical Observatory. This work was carried out in part at the Jet Propulsion Laboratory, which is operated for NASA by the California Institute of Technology. LB acknowledges support from CONICYT project Basal AFB-170002. C.W.L. is supported by the Basic Science Research Program through the National Research Foundation of Korea (NRF) funded by the Ministry of Education, Science and Technology (NRF-2019R1A2C1010851). This paper makes use of the following ALMA data: ADS/JAO.ALMA\#2019.1.00685.S. ALMA is a partnership of ESO (representing its member states), NSF (USA) and NINS (Japan), together with NRC (Canada), MOST and ASIAA (Taiwan), and KASI (Republic of Korea), in cooperation with the Republic of Chile. The Joint ALMA Observatory is operated by ESO, AUI/NRAO and NAOJ.

%%%%%%%%%%%%%%%%%%%%%%%%%%%%%%%%%%%%%%%%%%%%%%%%%%



%\bibliographystyle{mnras}


\onecolumn

\appendix

\section{}\label{app}

\begin{table*}
	\centering
	\caption{Selection of Compact objects identified in 3 mm continuum emission$^{a}$}
	\label{continuum}
	\begin{tabular}{cccccccccccc} % four columns, alignment for each
		\hline
IRAS  & RA &  DEC & $V_{\rm lsr}^b$  & Distance$^{c}$ &  $R_{\rm GC}^c$ &  ID  & Offset   & $r_{\rm aspect}$ & $R_{\rm eff}$ & $S_{\rm peak}$ & $S_{\rm total}$ \\
  &  & & (km~s$^{-1}$) & (kpc) & (kpc) & & ($\arcsec$,$\arcsec$)& & (pc)   & (Jy~beam$^{-1}$)  & (Jy) \\
\hline
I08076-3556   &   08:09:32.39  &  -36:05:13.2  & 5.9      & 0.4    &  8.45    & 1    &(8.56,15.49)  & 1.37  & 0.01  & 0.03  & 0.03  \\
              &                &               &          &        &          & 2    &(4.83,-4.90)  & 1.07  & 0.02  & 0.01  & 0.02  \\
I08303-4303   &   08:46:32.90  &  -43:13:54.0  & 14.3     & 2.3    &  8.96    & 1    &(3.87,7.99)  & 1.50  & 0.10  & 0.06  & 0.10  \\
I08448-4343   &   08:48:47.07  &  -43:54:35.9  & 3.7      & 0.7    &  8.43    & 1    &(2.86,-0.25)  & 3.86  & 0.05  & 0.04  & 0.09  \\
              &                &               &          &        &          & 2    &(21.53,7.53)  & 1.87  & 0.05  & 0.03  & 0.09  \\
I08470-4243   &   09:03:32.84  &  -42:54:31.0  & 12       & 2.1    &  8.83    & 1    &(8.15,4.60)  & 2.10  & 0.09  & 0.07  & 0.12  \\
I09002-4732   &   09:11:07.29  &  -47:44:00.8  & 3.1      & 1.2    &  8.45    & 1    &(0.32,-9.16)  & 1.08  & 0.03  & 3.74  & 4.26  \\
\hline
\multicolumn{12}{l}{$^a$ The full catalogue is available on line.}\\
\multicolumn{12}{l}{$^b$ The $V_{\rm lsr}$ values are from \citet{Bronfman1996}.}\\
\multicolumn{12}{l}{$^c$ The distances are from \citet{Urquhart2018} and \citet{Faundez2004}.}\\
	\end{tabular}
\end{table*}

\begin{table*}
	\centering
	\caption{Selection of compact objects identified in H$^{13}$CO$^{+} J=1-0 ^{a}$}
	\label{H13COP}
	\begin{tabular}{cccccccccc} % four columns, alignment for each
		\hline
IRAS   &  ID  & Offset   & $r_{\rm aspect}$ & $R_{\rm eff}$ & $d_{\rm peak}$ & $S_{\rm peak}$ & $S_{\rm total}$ & $V_{\rm cmp}$ & $\sigma$\\
  &   & ($\arcsec$,$\arcsec$)& & (pc) & (pc)  & (Jy~beam$^{-1}$~km~s$^{-1}$)  & (Jy~km~s$^{-1}$) & (km~s$^{-1}$) & (km~s$^{-1}$)\\
\hline
I08076-3556 &1 &(7.26   ,10.15  )& 3.39  & 0.06 & 0.01 & 3.12   &16.97    &6.21     & 0.28 \\
I08303-4303 &1 &(9.89   ,10.84  )& 1.57  & 0.26 & 0.07 & 8.71   &30.18    &14.53    & 1.00 \\
I08448-4343 &1 &(12.76  ,1.15   )& 3.16  & 0.09 & 0.03 & 5.37   &21.38    &2.78     & 0.84 \\
            &2 &(30.80  ,15.93  )& 1.77  & 0.07 & 0.11 & 3.62   &10.55    &5.06     & 0.66 \\
I08470-4243 &1 &(4.66   ,17.54  )& 1.16  & 0.22 & 0.14 & 4.06   &12.96    &13.38    & 0.94 \\
            &2 &(7.57   ,9.86   )& 3.86  & 0.23 & 0.05 & 3.78   &16.72    &12.06    & 0.74 \\
I09002-4732 &1 &(19.98  ,19.19  )& 1.70  & 0.10 & 0.20 & 8.48   &17.89    &2.44     & 1.08 \\
            &2 &(12.54  ,-11.26 )& 1.52  & 0.10 & 0.07 & 5.33   &11.92    &4.25     & 0.70 \\
            &3 &(18.09  ,-15.62 )& 2.74  & 0.09 & 0.11 & 4.33   &9.84     &2.54     & 0.62 \\
            &4 &(1.97   ,-36.41 )& 3.50  & 0.08 & 0.16 & 4.44   &8.91     &3.25     & 0.57 \\
\hline
\multicolumn{10}{l}{$^a$ The full catalogue is available on line.}\\
	\end{tabular}
\end{table*}

\begin{table*}
	\centering
	\caption{Selection of compact objects identified in H$^{13}$CN $J=1-0^{a}$}
	\label{H13CN}
	\begin{tabular}{cccccccc} % four columns, alignment for each
		\hline
IRAS   &  ID  & Offset   & $r_{\rm aspect}$ & $R_{\rm eff}$ & $d_{\rm peak}$ & $S_{\rm peak}$ & $S_{\rm total}$ \\
  &   & ($\arcsec$,$\arcsec$)& & (pc) & (pc)  & (Jy~beam$^{-1}$~km~s$^{-1}$)  & (Jy~km~s$^{-1}$) \\
\hline
I08076-3556 &1 &(9.76   ,14.74  )& 1.34  & 0.03 & 0.00 & 2.79   &6.29     \\
I08303-4303 &1 &(6.08   ,7.36   )& 1.19  & 0.14 & 0.03 & 17.00  &30.65    \\
I08448-4343 &1 &(1.27   ,0.58   )& 1.63  & 0.08 & 0.01 & 10.10  &33.88    \\
            &2 &(6.53   ,-13.19 )& 3.07  & 0.04 & 0.05 & 9.73   &18.57    \\
I08470-4243 &1 &(8.30   ,5.30   )& 3.36  & 0.15 & 0.01 & 13.50  &33.76    \\
I09002-4732 &1 &(24.39  ,21.85  )& 2.03  & 0.14 & 0.23 & 11.70  &41.69    \\
            &2 &(20.94  ,-12.69 )& 4.59  & 0.08 & 0.12 & 4.46   &9.71     \\
            &3 &(13.92  ,-10.52 )& 2.36  & 0.05 & 0.08 & 9.52   &13.40    \\
            &4 &(18.83  ,-23.57 )& 1.27  & 0.10 & 0.14 & 6.35   &14.80    \\
\hline
\multicolumn{8}{l}{$^a$ The full catalogue is available on line.}\\
	\end{tabular}
\end{table*}

\begin{table*}
	\centering
	\caption{Selection of compact objects identified in HCO$^{+}$ J=1-0$^{a}$}
	\label{HCOP}
	\begin{tabular}{cccccccccc} % four columns, alignment for each
		\hline
IRAS   &  ID  & Offset   & r$_{aspect}$ & R$_{eff}$ & d$_{peak}$ & S$_{peak}$ & S$_{total}$ & V$_{cmp}$ & $\sigma$\\
  &   & ($\arcsec$,$\arcsec$)& & (pc) & (pc)  & (Jy~beam$^{-1}$~km~s$^{-1}$)  & (Jy~km~s$^{-1}$) & (km~s$^{-1}$) & (km~s$^{-1}$)\\
\hline
\multicolumn{10}{c}{HCO$^{+}$ J=1-0}\\
\hline
I08076-3556 &1 &(10.31  ,13.96  )& 1.18  & 0.04 & 0.00 & 21.80  &57.20   &6.46     & 1.06 \\
I08303-4303 &1 &(11.51  ,6.91   )& 2.05  & 0.34 & 0.09 & 38.30  &182.70  &13.93    & 2.05 \\
I08448-4343 &1 &(4.88   ,1.10   )& 1.53  & 0.09 & 0.01 & 59.80  &247.60  &2.82     & 1.92 \\
            &2 &(24.45  ,11.80  )& 1.30  & 0.08 & 0.08 & 37.80  &127.00  &5.23     & 1.61 \\
I08470-4243 &1 &(7.60   ,6.26   )& 2.22  & 0.24 & 0.02 & 90.60  &325.40  &12.78    & 1.96 \\
I09002-4732 &1 &(18.83  ,14.75  )& 1.13  & 0.13 & 0.18 & 23.50  &73.10   &-0.69    & 0.22 \\
            &2 &(3.11   ,-2.59  )& 2.37  & 0.13 & 0.04 & 30.30  &104.70  &3.28     & 3.19 \\
            &3 &(14.81  ,-12.22 )& 1.16  & 0.13 & 0.09 & 33.70  &115.50  &5.57     & 1.45 \\
\hline
\multicolumn{8}{l}{$^a$ The full catalogue is available on line.}\\
	\end{tabular}
\end{table*}

\begin{table*}
	\centering
	\caption{Selection of compact objects identified in HCN $J=1-0$$^{a}$}
	\label{HCN}
	\begin{tabular}{cccccccc} % four columns, alignment for each
		\hline
IRAS   &  ID  & Offset   & $r_{\rm aspect}$ & $R_{\rm eff}$ & $d_{\rm peak}$ & $S_{\rm peak}$ & $S_{\rm total}$ \\
  &   & ($\arcsec$,$\arcsec$)& & (pc) & (pc)  & (Jy~beam$^{-1}$~km~s$^{-1}$)  & (Jy~km~s$^{-1}$) \\
\hline
I08076-3556 &1 &(8.32   ,15.53  )& 1.31  & 0.03 & 0.00 & 21.70  &45.16    \\
I08303-4303 &1 &(7.49   ,8.29   )& 1.25  & 0.19 & 0.04 & 78.50  &181.90   \\
I08448-4343 &1 &(3.31   ,0.00   )& 1.43  & 0.07 & 0.00 & 95.20  &272.50   \\
            &2 &(24.67  ,11.75  )& 1.15  & 0.06 & 0.08 & 48.70  &120.40   \\
I08470-4243 &1 &(8.56   ,3.96   )& 2.07  & 0.17 & 0.01 & 158.00 &404.00   \\
I09002-4732 &1 &(23.34  ,24.63  )& 3.26  & 0.16 & 0.24 & 34.50  &170.20   \\
            &2 &(17.77  ,-10.99 )& 2.25  & 0.13 & 0.10 & 46.40  &155.70   \\
            &3 &(12.29  ,-10.35 )& 2.37  & 0.09 & 0.07 & 42.80  &92.44    \\
            &4 &(15.72  ,-27.64 )& 1.73  & 0.17 & 0.14 & 32.70  &162.70   \\
\hline
\multicolumn{8}{l}{$^a$ The full catalogue is available on line.}\\
	\end{tabular}
\end{table*}

\begin{table*}
 \scriptsize
	\centering
	\caption{Parameters of selection from 119 sources$^a$}
	\label{LM}
	\begin{tabular}{ccccccccccc} % four columns, alignment for each
		\hline
IRAS   & R$_{\rm eff}$ & T$_d$ & log[L$_{bol}$]  & log[M$_{clump}$]  & log[L$'_{H^{13}CO^{+}}$] & log[L$'_{H^{13}CN}$] & log[L$'_{HCO^{+}}$] & log[L$'_{HCN}$] & log[M$_{vir}^{13}$] & log[M$_{vir}^{12}$]\\
      & (pc) & (K) & (L$_{\sun}$)  & (M$_{\sun}$)& (K~km~s$^{-1}$~pc$^{2}$)& (K~km~s$^{-1}$~pc$^{2}$) & (K~km~s$^{-1}$~pc$^{2}$) & (K~km~s$^{-1}$~pc$^{2}$)  & (M$_{\sun}$) & (M$_{\sun}$)\\
\hline
I08303-4303   &0.32 &  30 & 3.83 & 2.41 & -0.16 &-0.15  & 0.60  &0.62   & 2.33 & 3.07  \\
I08448-4343   &0.15 &  25 & 3.04 & 1.59 & -1.34 &-0.95  & -0.30 &-0.24  & 1.76 & 2.65  \\
I08470-4243   &0.32 &  33 & 4.04 & 2.36 & -0.25 &-0.19  & 0.77  &0.89   & 2.44 & 2.88  \\
I09002-4732   &0.24 &  39 & 4.59 & 2.41 & -1.13 &-1.08  & -0.21 &-0.24  & 1.58 & 3.12  \\
I09018-4816   &0.44 &  31 & 4.72 & 2.99 & -0.21 &0.11   & 0.80  &0.82   & 2.43 & 3.09  \\
\hline
\multicolumn{11}{l}{$^a$ The full catalogue is available on line.}\\
\multicolumn{11}{l}{$^b$ R$_{eff}$, T$_d$, L$_{bol}$  and M$_{clump}$ are compiled by \citet{Liu2020}, which are adopted from \citet{Urquhart2018} and \citet{Faundez2004}.}\\
	\end{tabular}
\end{table*}

\clearpage

\noindent
Author affiliations:\\

\noindent $^{1}$Key Laboratory for Research in Galaxies and Cosmology, Shanghai Astronomical Observatory, Chinese Academy of Sciences, 80 Nandan Road, Shanghai 200030, People’s Republic of China\\
$^{2}$Korea Astronomy and Space Science Institute, 776 Daedeokdaero, Yuseong-gu, Daejeon 34055, Republic of Korea\\
$^{3}$Department of Astronomy, The University of Texas at Austin, 2515 Speedway, Stop C1400, Austin, TX 78712-1205, USA\\
$^{4}$University of Science and Technology, Korea (UST), 217 Gajeong-ro, Yuseong-gu, Daejeon 34113, Republic of Korea\\
$^{5}$Jet Propulsion Laboratory, California Institute of Technology, 4800 Oak Grove Drive, Pasadena, CA 91109, USA\\
$^{6}$Institute of Astronomy and Astrophysics, Academia Sinica. 11F of Astronomy-Mathematics Building,
AS/NTU No. 1, Section 4, Roosevelt Rd., Taipei 10617, Taiwan\\
$^{7}$Center for Astrophysics $|$ Harvard \& Smithsonian, 60 Garden Street, Cambridge, MA 02138, USA\\
$^{8}$National Astronomical Observatory of Japan, National Institutes of Natural Sciences, 2-21-1 Osawa, Mitaka, Tokyo 181-8588, Japan\\
$^{9}$Kavli Institute for Astronomy and Astrophysics, Peking University, 5 Yiheyuan Road, Haidian District,
Beijing 100871, People's Republic of China\\
$^{10}$Department of Physics, P.O. Box 64, FI-00014, University of Helsinki, Finland\\
$^{11}$Departamento de Astronom\'{\i}a, Universidad de Chile, Las Condes, Santiago, Chile\\
$^{12}$School of Physics, University of New South Wales, Sydney, NSW 2052, Australia\\
$^{13}$School of Space Research, Kyung Hee University, Yongin-Si, Gyeonggi-Do 17104, Republic of Korea\\
$^{14}$National Astronomical Observatories, Chinese Academy of Sciences, Beijing, 100012, People's Republic of China \\
$^{15}$Key Laboratory of Radio Astronomy, Chinese Academy of Science, Nanjing 210008, People's Republic of China\\
$^{16}$Astronomy Department, University of California, Berkeley, CA 94720, USA\\
$^{17}$Department of Astronomy, Yunnan University, and Key Laboratory of Astroparticle Physics of Yunnan Province,
Kunming, 650091, People's Republic of China\\
$^{18}$IRAP, Universit\'{e} de Toulouse, CNRS, UPS, CNES, Toulouse, France\\
$^{19}$Indian Institute of Space Science and Technology, Thiruvananthapuram 695 547, Kerala, India\\
$^{20}$E\"{o}tv\"{o}s Lor\'{a}nd University, Department of Astronomy,
P\'{a}zm\'{a}ny P\'{e}ter s\'{e}t\'{a}ny 1/A, H-1117, Budapest, Hungary\\
$^{21}$Department of Astronomy, Peking University, 100871, Beijing, People's Republic of China\\
$^{22}$Departamento de Astronom\'{\i}a, Universidad de Concepci\'{o}n, Av. Esteban Iturra s/n, Distrito Universitario, 160-C, Chile\\
$^{23}$College of Science, Yunnan Agricultural University, Kunming, 650201, People's Republic of China\\
$^{24}$Shanghai Astronomical Observatory, Chinese Academy of Sciences, 80 Nandan Road, Shanghai 200030 People's Republic of China\\
$^{25}$Key Laboratory of Radio Astronomy, Chinese Academy of Sciences, Nanjing 210008, People's Republic of China\\
$^{26}$School of Physics and Astronomy, Sun Yat-sen University, 2 Daxue Road, Zhuhai, Guangdong, 519082, People's Republic of China

% Don't change these lines
\bsp	% typesetting comment
\label{lastpage}
\end{document}